%% file: paper.tex
\documentclass[letterpaper,twocolumn,10pt]{article}
\usepackage{usenix2019_v3}

\usepackage[english]{babel}
\usepackage{blindtext}

\usepackage{tikz}
\usepackage{pgfplots}
\usepackage{amsmath,amssymb,amsfonts}

\usepackage[utf8]{inputenc}
\usepackage[english]{babel}
\usepackage[T1]{fontenc}

\usepackage{mathtools}
\usepackage{algorithmic}
\usepackage[noend]{algorithm2e}
\usepackage{booktabs}
\usepackage{multirow}
\usepackage{paralist}
\usepackage{array, makecell}
\usepackage{boldline}
\usepackage{colortbl}
\usepackage{comment}
\usepackage{balance}
\usepackage{datetime}
\usepackage{enumitem}
\usepackage{color}
\usepackage{balance}
\usepackage{blindtext}
\usepackage{graphicx}
\usepackage{tcolorbox}
\usepackage{subcaption}
\usepackage[normalem]{ulem}
\usepackage{xspace}
\usepackage{color}
\usepackage{tabto}
\usepackage{siunitx}
\usepackage[export]{adjustbox}
\usepackage{collcell}
\usepackage{outlines}
\usepackage{newtxmath}
\usepackage{makecell}

\usepackage{filecontents}
\usepackage{cite}

\usepackage{pifont}%
\usepackage{xcolor}
\usepackage{titlesec}
\usepackage{dblfloatfix}
\usepackage{url}

\newcommand*{\MinNumber}{0.0}%
\newcommand*{\MidNumber}{0.99} %
\newcommand*{\MaxNumber}{2.0}%
\newcommand*{\TotMaxNumber}{7.28}%

\newcommand{\website}[0]{\texttt{https://bgproutes.io}}

\definecolor{applegreen}{rgb}{0.55, 0.71, 0.0}
\definecolor{chartreuse}{rgb}{0.5, 1.0, 0.0}
\definecolor{forestgreen}{rgb}{0.13, 0.55, 0.13}
\definecolor{lessThanOneYellow}{rgb}{1.0, 0.89, 0.0}
\definecolor{firstGreen}{rgb}{0.69, 1.0, 0.0}

\newcommand{\ApplyGradient}[1]{
        \setlength{\fboxsep}{3pt}%
        \ifdim #1 pt > \MidNumber pt
          \ifdim #1 pt > \MaxNumber pt
            \pgfmathsetmacro{\PercentColor}{max(min(100.0*(#1 - \MaxNumber)/(\TotMaxNumber-\MaxNumber),100.0),0.00)} %
            \hspace{-0.33em}\colorbox{forestgreen!\PercentColor!green}{#1}
          \else
            \pgfmathsetmacro{\PercentColor}{max(min(100.0*(#1 - \MidNumber)/(\MaxNumber-\MidNumber),100.0),0.00)} %
            \hspace{-0.33em}\colorbox{green!\PercentColor!firstGreen}{#1}
          \fi
        \else
          \pgfmathsetmacro{\PercentColor}{max(min(100.0*(\MidNumber - #1)/(\MidNumber-\MinNumber),100.0),0.00)} %
          \hspace{-0.33em}\colorbox{red!\PercentColor!lessThanOneYellow}{#1}
        \fi
}

\newcommand{\notimplies}{%
  \mathrel{{\ooalign{\hidewidth$\not\phantom{=}$\hidewidth\cr$\implies$}}}}

\newlength{\Oldarrayrulewidth}

\renewcommand{\arraystretch}{1.5}

\usetikzlibrary{arrows.meta}

\input{utils} %

\SetAlCapSty{} %

\definecolor{darkgreen}{rgb}{0.0, 0.5, 0.0}
\newcommand*\colourcheck[1]{%
  \expandafter\newcommand\csname #1check\endcsname{\textcolor{#1}{$\checkmark$}}%
}
\colourcheck{blue}
\colourcheck{green}
\colourcheck{red}
\colourcheck{darkgreen}

\definecolor{darkred}{rgb}{0.9, 0.17, 0.31}
\newcommand*\colourcross[1]{%
  \expandafter\newcommand\csname #1cross\endcsname{\textcolor{#1}{$\times$}}%
}
\colourcross{blue}
\colourcross{green}
\colourcross{red}
\colourcross{darkred}

\definecolor{mygray}{gray}{0.5}
\newcommand{\notesStart}[0]{\color{blue}}
\newcommand{\notesEnd}[0]{\color{black}}

\newcommand{\changedStart}[0]{\color{blue}}
\newcommand{\changedEnd}[0]{\color{black}}

\newcommand{\removedStart}[0]{\color{mygray}}
\newcommand{\removedEnd}[0]{\color{mygray}}

\excludecomment{removedblock}

\newcommand{\myitem}[1]{\vspace*{0.03in}\noindent\emph{\textbf{#1}}\enskip}

\newcommand{\remove}[1]{}

\newcommand{\name}[0]{\textsf{MVP}\xspace} %
\newcommand{\rname}[0]{\textsf{rMVP}\xspace}
\newcommand{\namevfour}[0]{\textsf{MVP$^{v4}$}\xspace} %
\newcommand{\namevsix}[0]{\textsf{MVP$_{v6}$}\xspace} %
\newcommand{\namevboth}[0]{\textsf{MVP$^{v4}_{v6}$}\xspace} %

\newcommand{\ris}[0]{{RIS}\xspace} %
\newcommand{\rv}[0]{{RV}\xspace} %
\newcommand{\spfa}[0]{\textit{reduction factor}\xspace}
\newcommand{\bb}[0]{{unbiased}\xspace}

\newtheorem{defn}{Definition}
\AfterEndEnvironment{defn}{\noindent\ignorespaces}

\def\BibTeX{{\rm B\kern-.05em{\sc i\kern-.025em b}\kern-.08em
    T\kern-.1667em\lower.7ex\hbox{E}\kern-.125emX}}

\makeatletter
\def\@copyrightspace{\relax}
\makeatother

\begin{document}

\title{Measuring Internet Routing from~the~\textcolor{cyan}{M}ost~\textcolor{cyan}{V}aluable~\textcolor{cyan}{P}oints}

\author{
{ \rm Thomas Alfroy\footnotemark[1],~~Thomas Holterbach\footnotemark[1],~~Thomas Krenc\footnotemark[2],~~KC Claffy\footnotemark[2],~~Cristel Pelsser\footnotemark[3]\vspace{-0.2cm}} \\
\footnotemark[1] University of Strasbourg, \footnotemark[2] CAIDA / UC San Diego, \footnotemark[3] UCLouvain\vspace{-.5cm} \\\\
\vspace{0.5cm} \textup{\textbf{\url{https://bgproutes.io}} \vspace{-1cm}}
} %

\maketitle

\input{abstract}

\input{introduction}

\input{background} %

\input{problem}
\input{opportunity}

\input{challenge}

\input{overview-kc}

\input{design}

\input{software}

\input{evaluation}

\input{soundness_design}

\input{related_work}

\input{conclusion}

\input{ethics.tex}
\input{ack.tex}

\bibliographystyle{plain}
\bibliography{reference}

\section*{Appendix}
\appendix

\input{survey}

\input{eval_extended}

\end{document}

%% file: utils.tex
\newcommand{\cref}[1]{Chapter~\ref{#1}}

\newcommand{\fref}[1]{Fig.~\ref{#1}}
\newcommand{\tref}[1]{Table~\ref{#1}}
\newcommand{\first}{\emph{(i)}\xspace}
\newcommand{\second}{\emph{(ii)}\xspace}
\newcommand{\third}{\emph{(iii)}\xspace}
\newcommand{\fourth}{\emph{(iv)}\xspace}

\newcommand{\First}{\emph{I}\xspace}
\newcommand{\Second}{\emph{II}\xspace}
\newcommand{\Third}{\emph{III}\xspace}
\newcommand{\Fourth}{\emph{IV}\xspace}
\newcommand{\Fifth}{\emph{V}\xspace}

\newcommand{\Firstp}{\emph{(I)}\xspace}
\newcommand{\Secondp}{\emph{(II)}\xspace}
\newcommand{\Thirdp}{\emph{(III)}\xspace}
\newcommand{\Fourthp}{\emph{(IV)}\xspace}

\newcommand{\ie}{i.e., \@}
\newcommand{\eg}{e.g., \@}

\newcommand{\etal}{et~al.\xspace}

\definecolor{darkgreen}{rgb}{0,0.5,0}
\definecolor{brown}{rgb}{0.7,0.3,0}
\definecolor{darkblue}{rgb}{0,0,0.5}
\definecolor{darkred}{rgb}{0.5,0,0}

\newcounter{fn1}
\newcounter{fn2}
\newcounter{fn3}
\newcounter{fn4}
\newcounter{fn5}
\newcommand{\link}{%
  \hspace{-0.2cm}
	\scalebox{0.75}{%
  \begin{tikzpicture}[baseline=+1.2ex]
    \draw [thick, outer sep=0.2] (-0.45,0.32) -- (-0.1,0.32);

	\end{tikzpicture}%
	}\hspace{-0.1cm}
}

\newcommand*\circledl[1]{\tikz[baseline=(char.base)]{
            \node[shape=circle,inner sep=1.6pt,fill=lightgray, scale=0.9] (char) {#1};}}

\newcommand*\frontaleye{%
\scalebox{0.25}{
\tikzset{every picture/.style={line width=0.75pt}}
\begin{tikzpicture}[x=0.75pt,y=0.75pt,yscale=-1,xscale=1]
\draw  [draw opacity=0][fill={rgb, 255:red, 0; green, 184; blue, 241 }  ,fill opacity=1 ] (104.5,179.99) .. controls (104.5,171.39) and (111.48,164.41) .. (120.08,164.41) .. controls (128.68,164.41) and (135.66,171.39) .. (135.66,179.99) .. controls (135.66,188.6) and (128.68,195.57) .. (120.08,195.57) .. controls (111.48,195.57) and (104.5,188.6) .. (104.5,179.99) -- cycle ;
\draw  [fill={rgb, 255:red, 0; green, 0; blue, 0 }  ,fill opacity=1 ] (84.83,179.7) .. controls (84.92,169.21) and (100.66,160.85) .. (119.99,161.01) .. controls (139.32,161.18) and (154.92,169.81) .. (154.83,180.3) .. controls (152.49,171.04) and (137.8,163.81) .. (119.97,163.66) .. controls (102.13,163.51) and (87.32,170.48) .. (84.83,179.7) -- cycle ;
\draw  [fill={rgb, 255:red, 0; green, 0; blue, 0 }  ,fill opacity=1 ] (113.6,179.91) .. controls (113.6,176.19) and (116.63,173.16) .. (120.36,173.16) .. controls (124.08,173.16) and (127.11,176.19) .. (127.11,179.91) .. controls (127.11,183.64) and (124.08,186.67) .. (120.36,186.67) .. controls (116.63,186.67) and (113.6,183.64) .. (113.6,179.91) -- cycle ;
\draw  [draw opacity=0][fill={rgb, 255:red, 255; green, 255; blue, 255 }  ,fill opacity=1 ] (195,141) .. controls (195,138.24) and (197.24,136) .. (200,136) .. controls (202.76,136) and (205,138.24) .. (205,141) .. controls (205,143.76) and (202.76,146) .. (200,146) .. controls (197.24,146) and (195,143.76) .. (195,141) -- cycle ;
\draw  [fill={rgb, 255:red, 0; green, 0; blue, 0 }  ,fill opacity=1 ] (154.65,179.7) .. controls (154.65,190.58) and (139.02,199.41) .. (119.74,199.41) .. controls (100.46,199.41) and (84.83,190.58) .. (84.83,179.7) .. controls (87.23,189.3) and (101.95,196.67) .. (119.74,196.67) .. controls (137.53,196.67) and (152.25,189.3) .. (154.65,179.7) -- cycle ;
\draw  [draw opacity=0][fill={rgb, 255:red, 255; green, 255; blue, 255 }  ,fill opacity=1 ] (125.45,172.24) .. controls (125.45,170.4) and (126.94,168.91) .. (128.78,168.91) .. controls (130.62,168.91) and (132.11,170.4) .. (132.11,172.24) .. controls (132.11,174.08) and (130.62,175.57) .. (128.78,175.57) .. controls (126.94,175.57) and (125.45,174.08) .. (125.45,172.24) -- cycle ;
\end{tikzpicture}
}\kern-.5em}

\newcommand*\frontaleyesmall{%
\scalebox{0.16}{
\tikzset{every picture/.style={line width=0.75pt}}
\begin{tikzpicture}[x=0.75pt,y=0.75pt,yscale=-1,xscale=1]
\draw  [draw opacity=0][fill={rgb, 255:red, 0; green, 184; blue, 241 }  ,fill opacity=1 ] (104.5,179.99) .. controls (104.5,171.39) and (111.48,164.41) .. (120.08,164.41) .. controls (128.68,164.41) and (135.66,171.39) .. (135.66,179.99) .. controls (135.66,188.6) and (128.68,195.57) .. (120.08,195.57) .. controls (111.48,195.57) and (104.5,188.6) .. (104.5,179.99) -- cycle ;
\draw  [fill={rgb, 255:red, 0; green, 0; blue, 0 }  ,fill opacity=1 ] (84.83,179.7) .. controls (84.92,169.21) and (100.66,160.85) .. (119.99,161.01) .. controls (139.32,161.18) and (154.92,169.81) .. (154.83,180.3) .. controls (152.49,171.04) and (137.8,163.81) .. (119.97,163.66) .. controls (102.13,163.51) and (87.32,170.48) .. (84.83,179.7) -- cycle ;
\draw  [fill={rgb, 255:red, 0; green, 0; blue, 0 }  ,fill opacity=1 ] (113.6,179.91) .. controls (113.6,176.19) and (116.63,173.16) .. (120.36,173.16) .. controls (124.08,173.16) and (127.11,176.19) .. (127.11,179.91) .. controls (127.11,183.64) and (124.08,186.67) .. (120.36,186.67) .. controls (116.63,186.67) and (113.6,183.64) .. (113.6,179.91) -- cycle ;
\draw  [draw opacity=0][fill={rgb, 255:red, 255; green, 255; blue, 255 }  ,fill opacity=1 ] (195,141) .. controls (195,138.24) and (197.24,136) .. (200,136) .. controls (202.76,136) and (205,138.24) .. (205,141) .. controls (205,143.76) and (202.76,146) .. (200,146) .. controls (197.24,146) and (195,143.76) .. (195,141) -- cycle ;
\draw  [fill={rgb, 255:red, 0; green, 0; blue, 0 }  ,fill opacity=1 ] (154.65,179.7) .. controls (154.65,190.58) and (139.02,199.41) .. (119.74,199.41) .. controls (100.46,199.41) and (84.83,190.58) .. (84.83,179.7) .. controls (87.23,189.3) and (101.95,196.67) .. (119.74,196.67) .. controls (137.53,196.67) and (152.25,189.3) .. (154.65,179.7) -- cycle ;
\draw  [draw opacity=0][fill={rgb, 255:red, 255; green, 255; blue, 255 }  ,fill opacity=1 ] (125.45,172.24) .. controls (125.45,170.4) and (126.94,168.91) .. (128.78,168.91) .. controls (130.62,168.91) and (132.11,170.4) .. (132.11,172.24) .. controls (132.11,174.08) and (130.62,175.57) .. (128.78,175.57) .. controls (126.94,175.57) and (125.45,174.08) .. (125.45,172.24) -- cycle ;
\end{tikzpicture}
}\kern-.5em}

\newcommand*\nofrontaleye{%
\scalebox{0.25}{
\tikzset{every picture/.style={line width=0.75pt}}
\begin{tikzpicture}[x=0.75pt,y=0.75pt,yscale=-1,xscale=1]
\draw  [draw opacity=0][fill={rgb, 255:red, 140; green, 196; blue, 74 }  ,fill opacity=0 ] (104.5,179.99) .. controls (104.5,171.39) and (111.48,164.41) .. (120.08,164.41) .. controls (128.68,164.41) and (135.66,171.39) .. (135.66,179.99) .. controls (135.66,188.6) and (128.68,195.57) .. (120.08,195.57) .. controls (111.48,195.57) and (104.5,188.6) .. (104.5,179.99) -- cycle ;
\draw  [fill={rgb, 255:red, 0; green, 0; blue, 0 }  ,fill opacity=0 ] (84.83,179.7) .. controls (84.92,169.21) and (100.66,160.85) .. (119.99,161.01) .. controls (139.32,161.18) and (154.92,169.81) .. (154.83,180.3) .. controls (152.49,171.04) and (137.8,163.81) .. (119.97,163.66) .. controls (102.13,163.51) and (87.32,170.48) .. (84.83,179.7) -- cycle ;
\draw  [fill={rgb, 255:red, 0; green, 0; blue, 0 }  ,fill opacity=0 ] (113.6,179.91) .. controls (113.6,176.19) and (116.63,173.16) .. (120.36,173.16) .. controls (124.08,173.16) and (127.11,176.19) .. (127.11,179.91) .. controls (127.11,183.64) and (124.08,186.67) .. (120.36,186.67) .. controls (116.63,186.67) and (113.6,183.64) .. (113.6,179.91) -- cycle ;
\draw  [draw opacity=0][fill={rgb, 255:red, 255; green, 255; blue, 255 }  ,fill opacity=0 ] (195,141) .. controls (195,138.24) and (197.24,136) .. (200,136) .. controls (202.76,136) and (205,138.24) .. (205,141) .. controls (205,143.76) and (202.76,146) .. (200,146) .. controls (197.24,146) and (195,143.76) .. (195,141) -- cycle ;
\draw  [fill={rgb, 255:red, 0; green, 0; blue, 0 }  ,fill opacity=0 ] (154.65,179.7) .. controls (154.65,190.58) and (139.02,199.41) .. (119.74,199.41) .. controls (100.46,199.41) and (84.83,190.58) .. (84.83,179.7) .. controls (87.23,189.3) and (101.95,196.67) .. (119.74,196.67) .. controls (137.53,196.67) and (152.25,189.3) .. (154.65,179.7) -- cycle ;
\draw  [draw opacity=0][fill={rgb, 255:red, 255; green, 255; blue, 255 }  ,fill opacity=0 ] (125.45,172.24) .. controls (125.45,170.4) and (126.94,168.91) .. (128.78,168.91) .. controls (130.62,168.91) and (132.11,170.4) .. (132.11,172.24) .. controls (132.11,174.08) and (130.62,175.57) .. (128.78,175.57) .. controls (126.94,175.57) and (125.45,174.08) .. (125.45,172.24) -- cycle ;
\end{tikzpicture}
}\kern-.5em}

\definecolor{green1}{RGB}{60, 225, 40}
\definecolor{green2}{RGB}{82, 206, 42}
\definecolor{green3}{RGB}{100, 180, 55}

\definecolor{blue1}{RGB}{126, 160, 242}

\definecolor{red1}{RGB}{0.07, 0.04, 0.56}
\definecolor{red2}{RGB}{0.07, 0.04, 0.56}
\definecolor{red3}{RGB}{0.07, 0.04, 0.56}

%% file: abstract.tex
\begin{abstract} 
    While the increasing number of Vantage Points (VPs) in RIPE RIS and RouteViews
    improves our understanding of the Internet, the quadratically increasing 
    volume of collected data poses a challenge to the scientific and operational
    use of the data.  The design and implementation of BGP and BGP data 
    collection systems lead to data archives with enormous redundancy, as
    there is substantial overlap in announced routes across many different VPs. 
    Researchers thus often resort to arbitrary sampling of the data, 
    which we demonstrate
    comes at a cost to the accuracy and coverage of previous works.  The continued
    growth of the Internet, and of these collection systems, exacerbates
    this cost.  The community needs a better approach to managing 
    and using these data archives.

    We propose \name, a system that 
	scores VPs according to their level of redundancy with other VPs,
	allowing more informed sampling of these data archives. 

    Our challenge is that the degree of redundancy between two updates depends
    on how we define {\em redundancy}, which in turn depends on
    the analysis objective. Our key contribution is
    a general framework and associated algorithms to assess 
   redundancy between VP observations. 
    We quantify the benefit of our approach for four 
canonical BGP routing analyses: AS relationship inference, AS rank computation, hijack detection, and routing detour detection.
\name improves the coverage or accuracy (or both) of all these analyses
while processing the same volume of data. 

\end{abstract}

%% file: introduction.tex
\section{Introduction}

\begin{figure*}[ht]
    \begin{minipage}{0.33\textwidth}
        \vspace{-0.5cm}
    \centering

    \begin{subfigure}[t]{.22\textwidth}
        \resizebox{1.15\linewidth}{!}{
            \hspace{-0.7cm}

            \begin{tikzpicture}
                \tikzset{vertex/.style = {shape=circle,minimum size=2em,fill=lightgray}}
                \tikzset{vertexeye/.style = {shape=circle,minimum size=2em,fill=white,opacity=0,scale=1.75}}

                \tikzset{edge1/.style = {line width=2pt, dotted}}
                \tikzset{edge2/.style = {-{Latex[length=3.2mm]},line width=2pt}}
            
                \node[vertex] (1) at  (0,0) {\LARGE \textbf{1}};
                \node[vertexeye] (0) at  (0,0.6) {\frontaleye};
                \node[vertex] (2) at  (2,0) {\LARGE  \textbf{2}};
                \node[vertexeye] (0) at  (2,0.6) {\nofrontaleye};

                \node[vertex] (3) at  (0,-2) {\LARGE  \textbf{3}};
                \node[vertex] (4) at  (2,-2) {\LARGE  \textbf{4}};
                \node[vertex] (5) at  (0,-4) {\LARGE  \textbf{5}};
                \node[vertex] (6) at  (2,-4) {\LARGE  \textbf{6}};
                \node[vertexeye] (0) at  (0,-4.4) {\nofrontaleye};
                \node[vertexeye] (0) at  (2,-4.4) {\nofrontaleye};
                \draw[edge1] (1) to (2);
                \draw[edge1, color=lightgray, opacity=0.5] (3) to (4);
                \draw[edge1, color=lightgray, opacity=0.5] (5) to (6);
                \draw[edge2] (3) to (1);
                \draw[edge2] (4) to (2);
                \draw[edge2] (5) to (3);
                \draw[edge2] (6) to (4);
            
            \end{tikzpicture}
        }        
        \vspace{-0.8cm}

        \caption{}
        \label{fig:ex2}
    \end{subfigure}
    \hfill
    \begin{subfigure}[t]{.22\textwidth}
        \resizebox{1.15\linewidth}{!}{
            \hspace{-0.7cm}

            \begin{tikzpicture}
                \tikzset{vertex/.style = {shape=circle,minimum size=2em,fill=lightgray}}
                \tikzset{vertexeye/.style = {shape=circle,minimum size=2em,fill=white,opacity=0,scale=1.75}}

                \tikzset{edge1/.style = {line width=2pt, dotted}}
                \tikzset{edge2/.style = {-{Latex[length=3.2mm]},line width=2pt}}
            
                \node[vertex] (1) at  (0,0) {\LARGE \textbf{1}};
                \node[vertexeye] (0) at  (0.1,0.6) {\frontaleye};
                \node[vertex] (2) at  (2,0) {\LARGE  \textbf{2}};
                \node[vertexeye] (0) at  (2.1,0.6) {\frontaleye};
                \node[vertex] (3) at  (0,-2) {\LARGE  \textbf{3}};
                \node[vertex] (4) at  (2,-2) {\LARGE  \textbf{4}};
                \node[vertex] (5) at  (0,-4) {\LARGE  \textbf{5}};
                \node[vertex] (6) at  (2,-4) {\LARGE  \textbf{6}};
                \node[vertexeye] (0) at  (0.1,-4.4) {\nofrontaleye};
                \node[vertexeye] (0) at  (2.1,-4.4) {\nofrontaleye};
                \draw[edge1] (1) to (2);
                \draw[edge1, color=lightgray, opacity=0.5] (3) to (4);
                \draw[edge1, color=lightgray, opacity=0.5] (5) to (6);
                \draw[edge2] (3) to (1);
                \draw[edge2] (4) to (2);
                \draw[edge2] (5) to (3);
                \draw[edge2] (6) to (4);
            
            \end{tikzpicture}
        }        
        \vspace{-0.8cm}

        \caption{}
        \label{fig:ex3}
    \end{subfigure}
    \hfill
    \begin{subfigure}[t]{.22\textwidth}
        \resizebox{1.15\linewidth}{!}{
            \hspace{-0.7cm}

            \begin{tikzpicture}
                \tikzset{vertex/.style = {shape=circle,minimum size=2em,fill=lightgray}}
                \tikzset{vertexeye/.style = {shape=circle,minimum size=2em,fill=white,opacity=0,scale=1.75}}

                \tikzset{edge1/.style = {line width=2pt, dotted}}
                \tikzset{edge2/.style = {-{Latex[length=3.2mm]},line width=2pt}}
            
                \node[vertex] (1) at  (0,0) {\LARGE \textbf{1}};
                \node[vertexeye] (0) at  (0.1,0.6) {\nofrontaleye};

                \node[vertex] (2) at  (2,0) {\LARGE  \textbf{2}};
                \node[vertexeye] (0) at  (2.1,0.6) {\nofrontaleye};

                \node[vertex] (3) at  (0,-2) {\LARGE  \textbf{3}};
                \node[vertex] (4) at  (2,-2) {\LARGE  \textbf{4}};
                \node[vertex] (5) at  (0,-4) {\LARGE  \textbf{5}};
                \node[vertex] (6) at  (2,-4) {\LARGE  \textbf{6}};
                \node[vertexeye] (0) at  (0.1,-4.4) {\frontaleye};
                \node[vertexeye] (0) at  (2.1,-4.4) {\nofrontaleye};
                \draw[edge1] (1) to (2);
                \draw[edge1] (3) to (4);
                \draw[edge1] (5) to (6);
                \draw[edge2] (3) to (1);
                \draw[edge2, color=lightgray] (4) to (2);
                \draw[edge2] (5) to (3);
                \draw[edge2, color=lightgray, opacity=0.5] (6) to (4);
            
            \end{tikzpicture}
        }        
        \vspace{-0.8cm}

        \caption{}
        \label{fig:ex4}
    \end{subfigure}
    \hfill
     \begin{subfigure}[t]{.22\textwidth}
        \resizebox{1.15\linewidth}{!}{
            \hspace{-0.7cm}

        \begin{tikzpicture}
            \tikzset{vertex/.style = {shape=circle,minimum size=2em,fill=lightgray}}
            \tikzset{vertexeye/.style = {shape=circle,minimum size=2em,fill=white,opacity=0,scale=1.75}}

            \tikzset{edge1/.style = {line width=2pt, dotted}}
            \tikzset{edge2/.style = {-{Latex[length=3.2mm]},line width=2pt}}
        
            \node[vertex] (1) at  (0,0) {\LARGE \textbf{1}};
            \node[vertexeye] (0) at  (0.1,0.6) {\nofrontaleye};

            \node[vertex] (2) at  (2,0) {\LARGE  \textbf{2}};
            \node[vertexeye] (0) at  (2.1,0.6) {\nofrontaleye};

            \node[vertex] (3) at  (0,-2) {\LARGE  \textbf{3}};
            \node[vertex] (4) at  (2,-2) {\LARGE  \textbf{4}};
            \node[vertex] (5) at  (0,-4) {\LARGE  \textbf{5}};
            \node[vertex] (6) at  (2,-4) {\LARGE  \textbf{6}};
            \node[vertexeye] (0) at  (0.1,-4.4) {\frontaleye};
            \node[vertexeye] (0) at  (2.1,-4.4) {\frontaleye};

            \draw[edge1] (1) to (2);
            \draw[edge1] (3) to (4);
            \draw[edge1] (5) to (6);
            \draw[edge2] (3) to (1);
            \draw[edge2] (4) to (2);
            \draw[edge2] (5) to (3);
            \draw[edge2] (6) to (4);
        
        \end{tikzpicture}
    }     
    \vspace{-0.8cm}

    \caption{}
    \label{fig:ex5}
     
    \end{subfigure}
    \caption[Example]{Combining local views can help
    to map the AS topology.
    Gray links are not visible from routes collected by VPs (\frontaleyesmall).} 

    \label{fig:partial_view}
    \end{minipage}
    \hfill
    \begin{minipage}{0.63\textwidth}
        \centering
        \begin{subfigure}[t]{.31\textwidth}
            \captionsetup{justification=centering}

            \includegraphics[width=\textwidth]{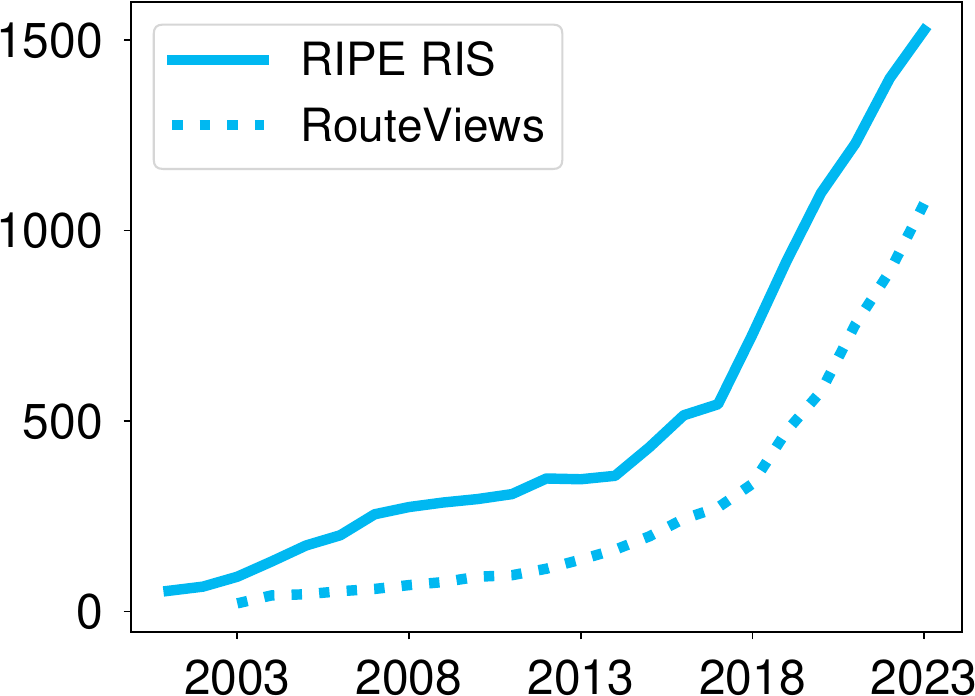}
            \vspace*{-3mm}
            \caption{Growth in VPs}
            \label{fig:ris_nb_vp2}
        \end{subfigure}
        \begin{subfigure}[t]{.31\textwidth}
            \captionsetup{justification=centering}

            \includegraphics[width=\textwidth]{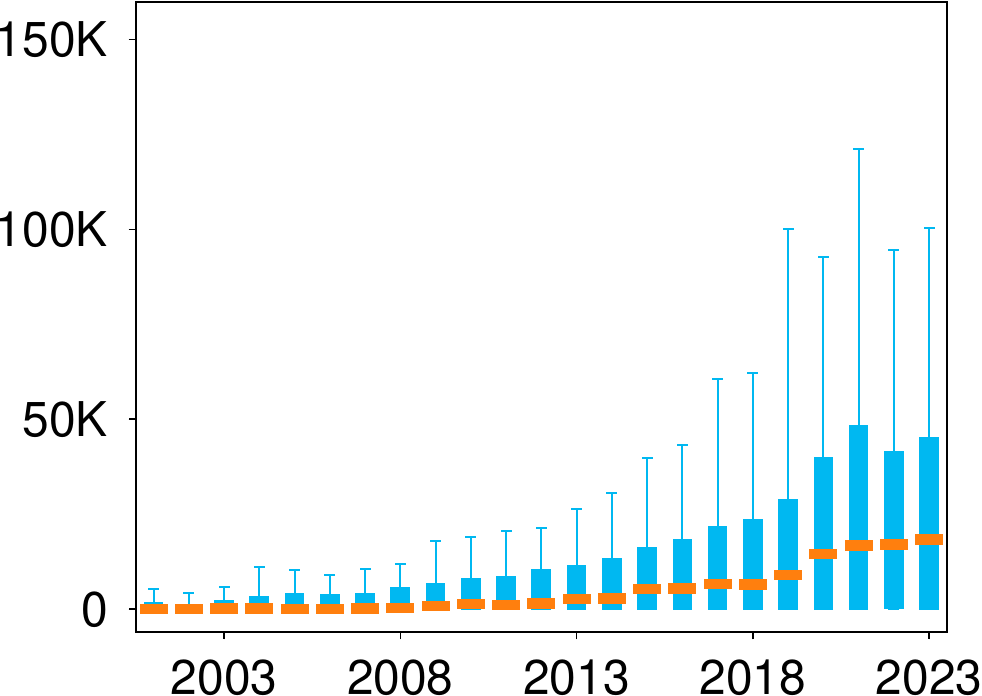}
            \vspace*{-3mm}
            \caption{Number of updates\\per VP and per hour.}
            \label{fig:update_vp2}
        \end{subfigure}
        \begin{subfigure}[t]{.31\textwidth}
            \captionsetup{justification=centering}

            \includegraphics[width=\textwidth]{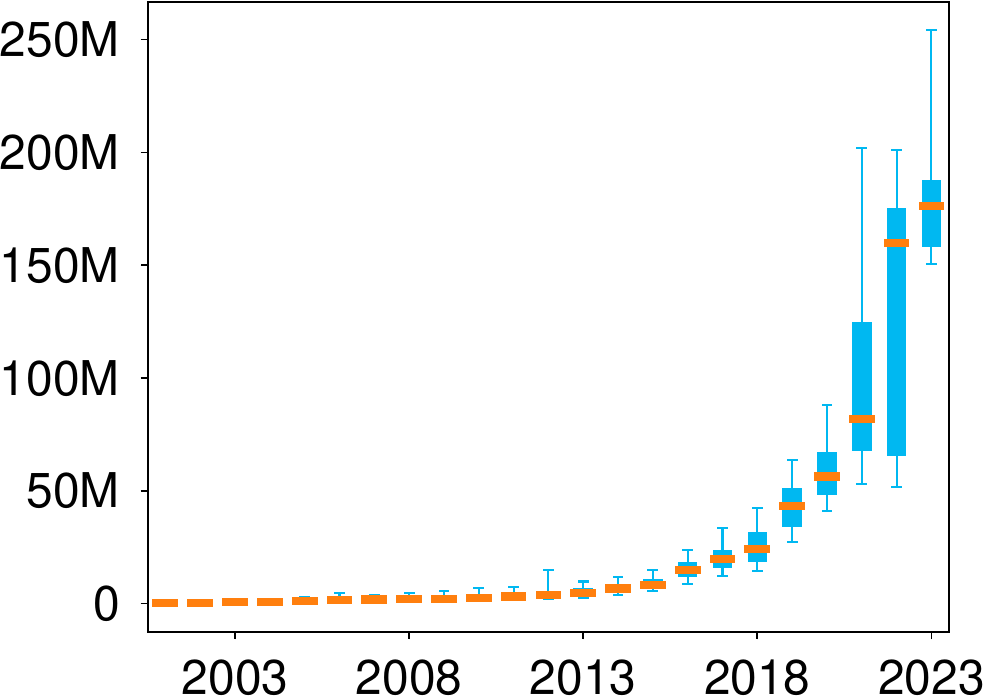}
            \vspace*{-3mm}
            \caption{Total number of updates per hour.}
            \label{fig:ris_nb_updates}
        \end{subfigure}
        \vspace*{-2mm}

        \caption{The number of VPs increases over time and so does the number of collected updates.
        Both \ris and \rv are considered in \fref{fig:update_vp2} and \ref{fig:ris_nb_updates}.}
        \label{fig:load}
    \end{minipage}
\end{figure*}

Routing information services such as RIPE RIS~\cite{RIPE} and RouteViews (RV)~\cite{RouteViews}
continuously collect and publish data from more than 2500 Vantage Points (VPs), 
each of which is a
BGP router that exports its best routes to the collection platform.
These data collection systems are critical to scientific 
as well as operational analyses of the global Internet infrastructure.
But these systems face a cost-benefit trade-off~\cite{aben_concern}. 
The information-hiding character of BGP means that improving the
visibility of the Internet routing system requires cultivating
many peering relationships with operators willing to contribute VPs
to the platform.  However,
deployment of new VPs amplifies the data management
requirements caused by the growth of the Internet itself:
the number of unique IP prefixes
(\eg due to de-aggregation or new
assignments)
constantly grows~\cite{cidr_report}, as well as the number of
unique ASes and links between them.  Even with a constant number of VPs,
the volume of routing data inevitably increases, contributing to a
quadratic increase of observed updates over time (\fref{fig:ris_nb_vp2}).
The situation presents a challenge for users, who often cannot or do not
want to process terabytes of (redundant) data.
Users often resort to sampling the data in arbitrary ways, 
such as grabbing all VPs on a single collector.

We design and implement a framework to optimize the use of these data collection 
systems, which will also lower the barrier to their use in 
lower-resourced circumstances.
Our design relies on the principle of {\em redundancy} in BGP data,
a delicate concept since even two identical updates from 
two different VPs may not be redundant (depending on the use case).
We take a deep dive into a context-specific framework for 
quantifying redundancy in BGP data,
grounded in operational principles and research use cases.
Our resulting system identifies a set of VPs whose exported
routes collectively exhibit a low level of redundancy—--enabling users
to prioritize the processing of the most valuable BGP updates.

\myitem{Contributions.}
We make the following contributions.

\begin{itemize}[leftmargin=*]
    \setlength{\itemsep}{4pt}

\item We perform a comprehensive analysis based on simulations 
and a survey that demonstrates the cost-benefit tradeoff of setting up new VPs,
and the value of strategically selecting them to analyze Internet routing.
We show that current approaches used by researchers to select VPs are
largely unoptimized, sacrificing coverage and accuracy of a wide range
of measurement studies and tools (\S\ref{sec:background}-\S\ref{sec:problems}). 

\item We characterize redundancy between
updates collected by different VPs. 
We explore different definitions of redundancy and find that
optimizing our algorithms for a given definition leads
to a undesirable overfitting effect (\S\ref{sec:opportunity}-\S\ref{sec:challenges}).

\item We design a system, \name, that returns a list of the ``most valuable'' VPs,
i.e., those that %
enable users to minimize data redundancy (regardless of how we define it) and prioritize valuable route updates.
\name relies on new data-driven algorithms that quantify redundancy between VPs
based on the four main BGP attributes (time, prefix, AS path, and communities)
while being robust against typical biases observed in the
Internet routing ecosystem (\S\ref{sec:solution_overview}-\S\ref{sec:design}).

\item
We run \name as a service at \website.
We benchmark \name and show that it optimizes (without overfitting)
the tradeoff between the volume of data used 
and its utility for many objectives (\S\ref{sec:functions}-\S\ref{sec:eval}).

\end{itemize}

\myitem{Impact on scientific measurement studies.}
The value of \name is its \textit{wide} impact.
Besides enabling a more systematic sampling of the \ris and \rv data archives,
it can consistently, and at no cost for users,
improve the accuracy and coverage
of measurement studies as well as monitoring tools fueled by BGP routes collected by
\ris and \rv.
To measure the impact of \name, we replicated the algorithms used in four studies/tools and used \name to select
the VPs from which they process BGP routes.
In all four cases, using \name improved the accuracy and coverage 
while processing the same data volume.
We inferred more AS relationships (+15\%), fixed errors in the AS rank dataset,
observed more routing detours (+44\%) while characterizing them more accurately,
and inferred more forged-origin hijacks (+35\%) with $\approx$4$\times$ less incorrect inferences (\ie false positives).

%% file: background.tex
\section{Background}
\label{sec:background}

RIPE's Routing Information Service (\ris)~\cite{RIPE} and RouteViews (\rv)~\cite{RouteViews}
are two widely-used platforms that collect 
BGP routes and make them available to the community.
These platforms use BGP speakers (\textit{a.k.a}. collectors) to peer with
BGP routers in order to collect routes exported by those routers.
We call \textbf{vantage points (VPs)}
the BGP routers that export their routes to a collection platform.
As of May 2023, 32\% of the \ris and \rv VPs 
~\cite{RIPE_peers, RouteViews_peers} are 
\textit{full feeders}, \ie they send a route for roughly all of the 
announced IP prefixes on the Internet ($\approx$941k prefixes~\cite{cidr_report}).
A BGP route mainly carries
routing information in four of its attributes \cite{bgpstream}: \first the timestamp 
at which the route was received, \second the IP (v4 or v6) prefix that the route announces,
\third the AS path used to reach that prefix, and \fourth a set
of BGP communities. Among other uses, researchers leverage the timestamp 
to find transient paths~\cite{transientConv},
the prefix to detect hijacks~\cite{artemis}, 
the AS paths to infer AS relationships~\cite{Luckie2013},
and the communities to measure unnecessary BGP traffic~\cite{unnecessary_updates}.

Each VP provides its \textit{local view}, i.e., only the BGP routes it observes. 
\fref{fig:partial_view} illustrates the effect of combining local views
for inferring the AS topology from the AS paths in BGP routes.
In \fref{fig:partial_view}, every AS runs a single BGP router,
owns one prefix, and announces it in BGP.
We configure 
routing policies based on the Gao-Rexford model~\cite{GaoRexford}, \ie routing paths follow a valley-free pattern.
Straight (resp. dashed) lines are customer-to-provider (resp. peer-to-peer) links.
With the local view of \circledl{1},  
one can infer all the AS links but
the two peering links
\mbox{\circledl{3} \link \circledl{4}} and
\mbox{\circledl{5} \link \circledl{6}} (\fref{fig:ex2}).
Combining the local views of \circledl{1} and \circledl{2} does not help to discover more links
(\fref{fig:ex3}).
With the local view of \circledl{5},
one can infer all the AS links but the two customer-to-provider links
\mbox{\circledl{2} \link \circledl{4}},
\mbox{\circledl{4} \link \circledl{6}} (\fref{fig:ex4}). Combining the local views 
of \circledl{5} and \circledl{6} enables discovery of the full topology (\fref{fig:ex5}).
However, observe that this last scenario is unlikely 
in practice as the location of the VPs is skewed with many more VPs
present in highly-connected or central (\eg Tier1) ASes~\cite{RIS_pavlos}.
Observe also that VPs can have a redundant view over the AS topology, \eg
the two VPs in \fref{fig:ex3} observe the same set of links.

By May 2023, \ris had 1526 VPs and \rv had 1071 VPs, and their number keeps increasing (\fref{fig:ris_nb_vp2}).
Users can download BGP routes exported by these platforms at the granularity of the VP (with some limitations~\cite{ripe_perpeer})
or the collector.
Users can download a RIB dump, i.e., a snapshot of the BGP routes seen 
by a VP at a particular time, which (in Jan. 2023) yielded $\approx$941k 
routes for a full feeder. Alternatively, users may download every single 
BGP update observed by the VPs
over time (\eg using~\cite{bgpstream}), which currently results in
$\approx$18K updates per hour (median in May 2023)
for a single VP (\fref{fig:update_vp2}), and billions of updates per
day for all \ris and \rv VPs (\fref{fig:ris_nb_updates}).

%% file: problem.tex
\section{Problem}
\label{sec:problems}

Deploying more VPs expands the visibility of the routing system
(\S\ref{sec:case1}), but also increases collected data
volumes raising barriers to its use (\S\ref{sec:problem2}).
We survey researchers and find that they resort to unoptimized sampling,
which they acknowledge can negatively impact the quality of their results (\S\ref{sec:problem3}).

\vspace{-0.15cm}
\subsection{More VPs improves data completeness}\hfill%
\label{sec:case1}

A tiny fraction (1.3\%) of the 74k ASes participating in the global
routing system~\cite{cidr_report} host a VP.
This fraction remains low (8.4\%) even when focusing on the 11441 transit ASes (\ie 
those with at least one customer).  
While we cannot know how much additional topology we might observe from VPs that do not
peer with the public collection systems,
we can estimate this gap using simulations of topologies
whose statistical parameters match those of the known global Internet.

\myitem{Methodology.}
We created a mini-Internet with 600 ASes, each running a single BGP
router.  We generated the AS topology using the Hyperbolic Graph
Generator~\cite{ALDECOA2015492}. We set the average node degree to 6.1,
which results in a comparable degree of connectivity (\textit{a.k.a.}
Beta index) to the one observed in CAIDA's AS relationship dataset
from December 2022~\cite{CAIDA_relationships}, and use as the 
degree distribution
a power law with exponent 2.1 (as in~\cite{ALDECOA2015492}).  We defined
the AS relationships as follows. The three ASes with the highest degree
are Tier1 ASes and are fully meshed. ASes directly connected to a Tier1
are Tier2s.  ASes directly connected to a Tier2 but not to a Tier1 are
Tier3s, etc. Two connected ASes have a peer-to-peer (p2p) relationship
if they are on the same level, and a customer-to-provider (c2p) relationship
if not.  The routing policies follow the Gao-Rexford model~\cite{GaoRexford}.

\begin{figure}[t]
    \centering
    \includegraphics[width=1\linewidth]{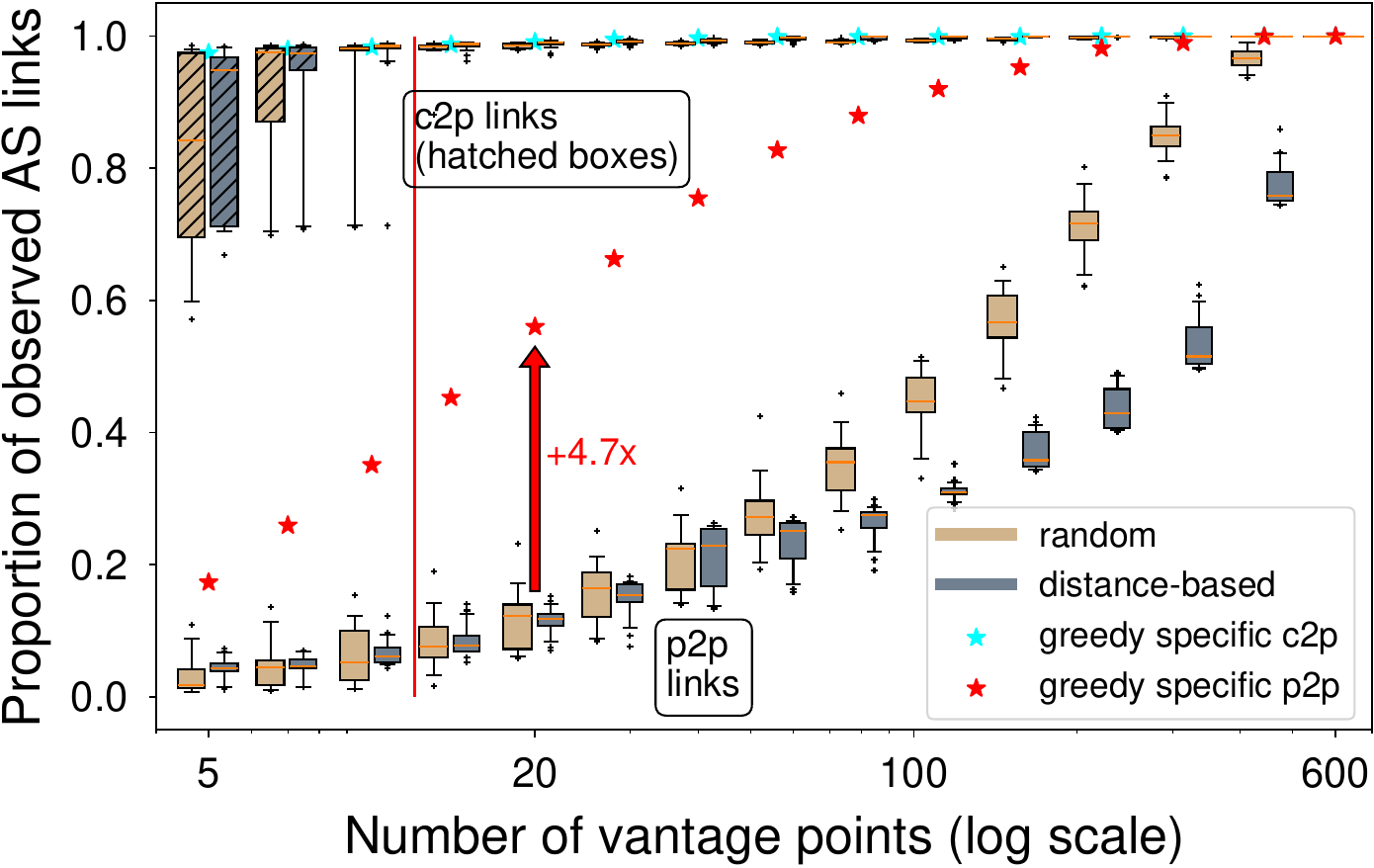}
    \caption{Simulations of a mini Internet with 600 ASes. We make two key observations:
    \first deploying more VPs helps to reveal more AS links, and \second
    arbitrarily selecting VPs performs poorly compared to selecting them with \textit{greedy specific} (a best-case approximation).
    The line in a box depicts the median value; the whiskers show the 5 and the 95th percentile.
}
    \label{fig:mini-internet_exp}
\end{figure}%

\fref{fig:mini-internet_exp} shows the proportion of observed AS links
as a function of the number of ASes hosting a VP. We consider 
three VP deployment strategies: \first \textit{random}, which randomly deploys VPs across all the ASes;
\second \textit{distance-based}, which aims to maximize the AS-level distance between the deployed VPs; and \third
\textit{greedy specific}, which approximates the best case for topology discovery using a greedy approach.
We ran every selection strategy twenty times (with different random seeds).
We computed the proportion of observed links and show separately the 
p2p and 
c2p links in \fref{fig:mini-internet_exp}. 

\myitem{Conclusions.}
Although we take the results with a grain of salt
because the topology differs (but exhibits similar patterns) from the visible portion of the actual (unknown) AS topology, we tentatively draw the following four conclusions.

\noindent
\Firstp As expected, for a given VP deployment strategy, more VPs often lead to more
links observed; all links are observed only when all ASes host a VP.

\noindent
\Secondp P2p  links are harder to observe than c2p links.
We find that p2p links are more visible from VPs at the edge.
This result is consistent with the fact that 
p2p links are generally not
advertised upwards in the Internet hierarchy when routing policies follow the Gao-Rexford model.

\noindent
\Thirdp The \textit{distance-based} deployment strategy performs poorly (even worse than random)
because it overprioritizes isolated VPs at the edge over some other important VPs in the core.

\noindent
\Fourthp When 1.3\% of ASes host a VP (same proportion 
as current \ris and \rv VPs), %
only $\approx$5\% of the p2p links are seen when using the \textit{random} deployment strategy.

\myitem{Confirmation with real (but private) data.} We contacted a private BGP
data provider (bgp.tools) that collects BGP routes from $\approx$1000 routers and 
compared the set of AS links observed from these private feeds against the set of AS links 
observed by \ris and \rv VPs (in September 2023). We find that 
the private data provider saw 192k AS links that \textit{none} of the \ris and \rv VPs observed, 
and vice versa, \ris and \rv VPs observed 401k links that the private data provider did not observe.
In either case, the lack of VPs leads to missing routing information. We can
thus expect---and hope for---the number of VPs to keep increasing.

\vspace{-0.15cm}
\subsection{BGP data management is challenging}\hfill%
\label{sec:problem2}

Deploying more VPs generates more data as each of them collects BGP updates.
Moreover, new IP prefixes advertised in BGP (see~\cite{cidr_report})
increase the volume of data collected by every VP as
it triggers the propagation of
new BGP routes that many VPs (\eg the full feeders) observe
and send to the collection platforms.
The compound effect---more VPs (\fref{fig:ris_nb_vp2})
and more updates per VP (\fref{fig:update_vp2})---yields 
a \textit{quadratic} increase in updates reaching the 
collection platforms (\fref{fig:ris_nb_updates}),
which challenges both users and data providers~\cite{aben_concern}.
Although several tools can speed up data processing~\cite{bgpstream,bgpkit,9464008},
many measurement studies and monitoring tools use only a
sample of data collected by \ris and \rv, either using only a subset of the VPs
or a short time window, or both\footnote{We purposively do not cite any paper
to preserve the anonymity of the respondents of our survey.}.
While authors do not typically explain why they do not use all the data,
the sampling suggests two (inter-related) explanations: authors believe the
sample is representative and sufficiently complete; and/or the data volume is
not worth trying to manage.  We confirm these explanations with a survey
that we conducted involving authors of eleven research papers.

\myitem{Methodology of our survey.} We classified eleven BGP-based studies from top
conferences\footnote{SIGCOMM, NSDI, S\&P, USENIX Security, NDSS and IMC. }
into two categories based on how they used BGP data.\footnote{A
	paper may be in both categories.}
Nine papers used \textbf{all} routes collected from 
a \textbf{subset} of the VPs (category $C_1$);  
six papers used a \textbf{short time frame} ($C_2$).
For each paper, we asked authors questions regarding
their use of BGP data: whether data volume limited their work, 
how and why they sampled BGP data sources, their understanding of
the impact on the quality of their results, and if they would do 
things differently if they had more resources or time.  
We did not receive answers from the authors of three papers.
Thus, we have seven respondents in $C_1$ and five in $C_2$.
We summarize the results here; details of the survey are in
an appendix (\S\ref{sec:survey}).

\myitem{The volume of BGP data to process is often a limiting factor.}
Seven (of eight) respondents found the BGP data expensive to process.
For three respondents in $C_1$, processing time motivated them
to use only a subset of the VPs; three respondents in $C_2$ 
considered the processing time when choosing a measurement interval.
Even a respondent who used a Spark cluster found it 
inhibitively time-consuming to process the BGP data. 

\myitem{Respondents in $C_1$ selected VPs in an unoptimized fashion.}
One respondent picked geographically distant BGP collectors.
Our experiments
(\fref{fig:mini-internet_exp}) and evaluation (\S\ref{sec:eval})
show that this strategy, while intuitive, often fails to optimize for any given
metric (\eg coverage).  Other respondents said they chose VPs randomly,
or those with the highest number of prefixes.  Another responded to have
unintentionally discarded some VPs, leaving an arbitrarily selected set
in the study.  Two respondents did not remember how they selected VPs.

\vspace{-0.15cm}
\subsection{Unoptimized sampling negatively impacts the quality of the results}
\label{sec:problem3}

We show the negative effects of an unoptimized sampling using 
our controlled simulations as well as our survey.

\myitem{Selecting VPs arbitrarily performs poorly.}
Our mini-Internet simulation (\S\ref{sec:case1})
showed that arbitrary VP selection strategies
perform significantly worse than \textit{greedy specific} (a best-case 
approximation) when the goal is to map the AS topology.
For instance, randomly selecting 20 VPs reveals 12\% of the p2p links compared 
to 56\% when selecting them using \textit{greedy specific}---a 4.7$\times$ improvement factor that we highlight in \fref{fig:mini-internet_exp}.
Our evaluation reveals that this performance gap between using an arbitrary VPs selection
strategy and a best-case approach also exists for various other metrics,
\eg hijacks or transient paths detection (\S\ref{sec:eval}).

\myitem{Six respondents in $C_1$ acknowledged that using more VPs would 
improve the quality of their analysis.}
The last respondent was not sure, given the potential redundancy in
the data sources (which he did not analyze). 
Two of the six believed it would not {\em significantly} change 
the conclusion of their measurement studies
(\eg one said that it could help to pinpoint corner cases).
However, six of the seven authors in $C_1$ affirmed that they would have
used more VPs if they had more resources and time.

\myitem{All five respondents in $C_2$ said that extending the duration of
their study would improve the quality of their results.}
One respondent thought the gain would not be significant;
another said it could help detect rare routing events. 
All respondents in $C_2$ would have extended the duration 
of their observation window given more time and resources. 
We experimentally confirm in \S\ref{sec:eval_detours} that extending 
the timeframe of analysis improves the quality of its results 
with a case study on routing detour characterization~\cite{forwarding-detour}.

%% file: opportunity.tex
\section{Opportunity to Optimize Sampling}
\label{sec:opportunity}

We propose a systematic framework to characterize
{\em redundancy} across BGP routes collected by the VPs.
We use the term {\em redundant} to refer to updates with 
similar (or identical, depending on the \textit{redundancy} definition)
attribute values (see definitions below).
Thus, two redundant VPs, \ie that observe
redundant routes, likely provide similar views over
routing events such as hijacks, traffic engineering, etc.

\myitem{Methodology.}
We characterize redundancy between pairs of VPs by computing the proportion of redundant updates 
that they collect using three different, gradually stricter, 
definitions of update redundancy.
We denote $U_i$ the set of updates observed by VP $i$.
Consider a BGP update $u_{t,p} \in U_i$
with $t$ the time at which the route was observed and $p$ its prefix.
\begin{defn}[prefix based]
    The update $u_{t_1,p_1} \in U_1$ is redundant with the update $u_{t_2,p_2} \in U_2$ if:
    \begin{itemize}
        \setlength\itemsep{-0.1cm}
        \item $\lvert t_1 - t_2 \lvert < 5$ minutes, and $p_1 = p_2$.
   \end{itemize}
    \label{def:red1}
  \end{defn}
\noindent
We chose 5 minutes because it is an approximation of the BGP convergence time~\cite{transientConv}. This first
definition might be appropriate to map prefixes with their origin AS.

For our second definition, we denote $A_i(t,p)$ the set of AS links in the AS path of the most recent BGP route observed by 
VP $i$ for prefix $p$ at time $t$.
\begin{defn}[prefix and as-path based]
    The update $u_{t_1,p_1}$ $\in U_1$ is redundant with the update $u_{t_2,p_2} \in U_2$ if:
    \begin{itemize}
        \setlength\itemsep{-0.1cm}
        \item $\lvert t_1 - t_2 \lvert < 5$ minutes, and $p_1 = p_2$, and
        \item $A_1(t_1, p_1) \setminus A_1(t_1-\epsilon, p_1) \subset A_2(t_2, p_2) \setminus A_2(t_2-\epsilon, p_2)$.
   \end{itemize}
   \label{def:red2}
\end{defn}
\noindent
The second condition checks whether the changes (operator $\setminus$)
in the AS paths observed by VP $1$ for a given prefix are included (operator $\subset$) 
in the set of changes observed by VP $2$ for the same prefix. This second definition
might be appropriate to detect new AS links or transient paths.

Our third definition 
follows the same approach but adds BGP communities.
We denote $C_i(t,p)$ the set of community values of the most recent BGP route observed by 
VP $i$ for prefix $p$ and at time $t$.
\noindent
\begin{defn}[prefix, as-path, and community-based]
    The update $u_{t_1,p_1} \in U_1$ is redundant with update $u_{t_2,p_2} \in U_2$ if:
    \begin{itemize}
        \setlength\itemsep{-0.1cm}
        \item $\lvert t_1 - t_2 \lvert < 5$ minutes, and $p_1 = p_2$, and
        \item $A_1(t_1, p_1) \setminus A_1(t_1-\epsilon, p_1) \subset A_2(t_2, p_2) \setminus A_2(t_2-\epsilon, p_2)$, and
        \item $C_1(t_1, p_1) \setminus C_1(t_1-\epsilon, p_1) \subset C_2(t_2, p_2) \setminus C_2(t_2-\epsilon, p_2)$.
   \end{itemize}
   \label{def:red3}
\end{defn}
We note that Def. \ref{def:red2} and \ref{def:red3} are asymmetric because, given two set $X$ 
and $Y$ of objects of same type, $X \subset Y \notimplies Y \subset X$.

\begin{figure}[t!] %

    \begin{subfigure}{0.05\textwidth}

        \includegraphics[width=\textwidth]{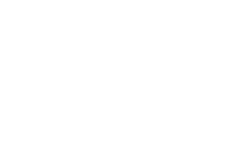}
        
    \end{subfigure}
    \begin{subfigure}{0.38\textwidth}

        \includegraphics[width=\textwidth]{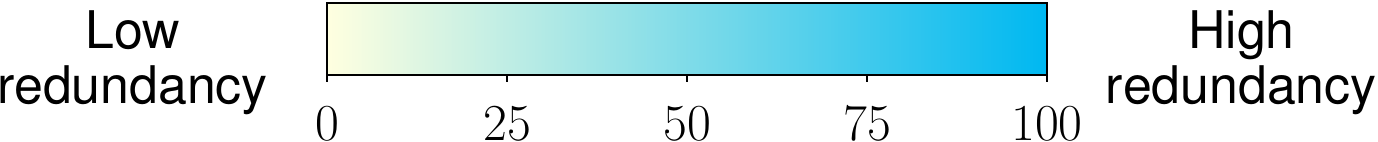}

    \end{subfigure}
    \hspace{0.5cm}
        
    \medskip

    \begin{subfigure}{0.013\textwidth}

        \includegraphics[width=\textwidth]{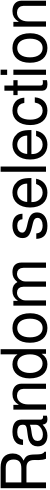}
        
    \end{subfigure}
    \begin{subfigure}{0.000\textwidth}

        \includegraphics[width=\textwidth]{figures/white.png}
        
    \end{subfigure}\hspace*{\fill}
    \begin{subfigure}{0.145\textwidth}

    \includegraphics[width=\textwidth]{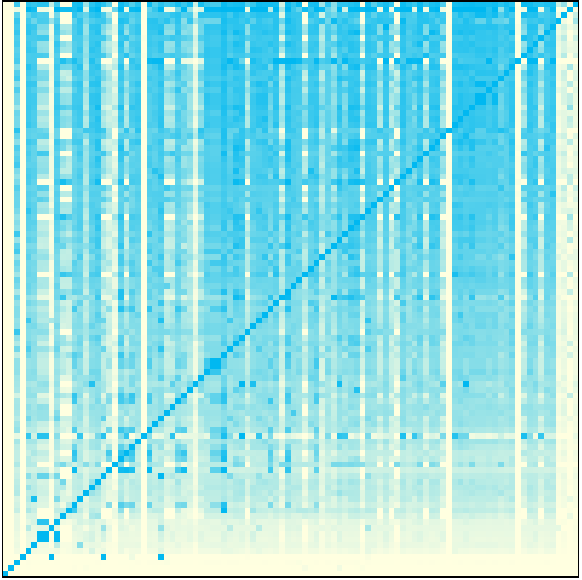}
    
    \end{subfigure}\hspace*{\fill}
    \begin{subfigure}{0.145\textwidth}

        \includegraphics[width=\textwidth]{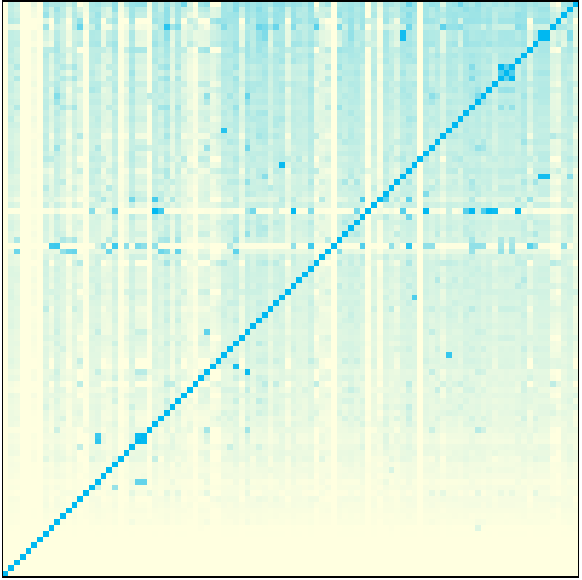}

    \end{subfigure}\hspace*{\fill}
    \begin{subfigure}{0.145\textwidth}

        \includegraphics[width=\textwidth]{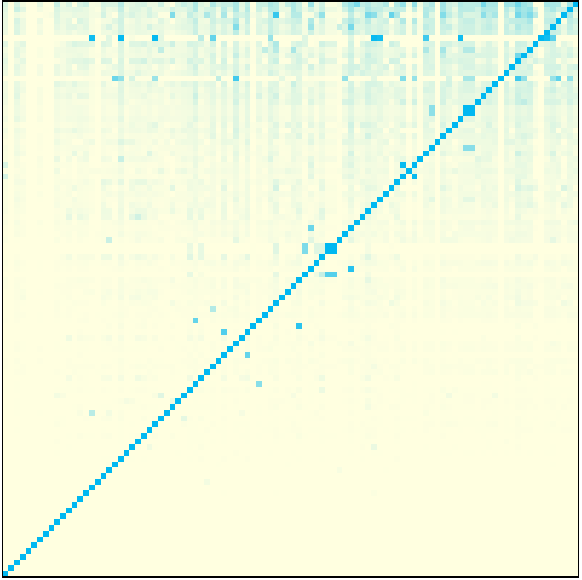}

    \end{subfigure}

    \medskip
    \begin{subfigure}{0.017\textwidth}

        \includegraphics[width=\textwidth]{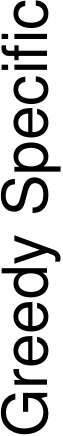}
        
    \end{subfigure}\hspace*{\fill}
    \begin{subfigure}{0.145\textwidth}

        \includegraphics[width=\textwidth]{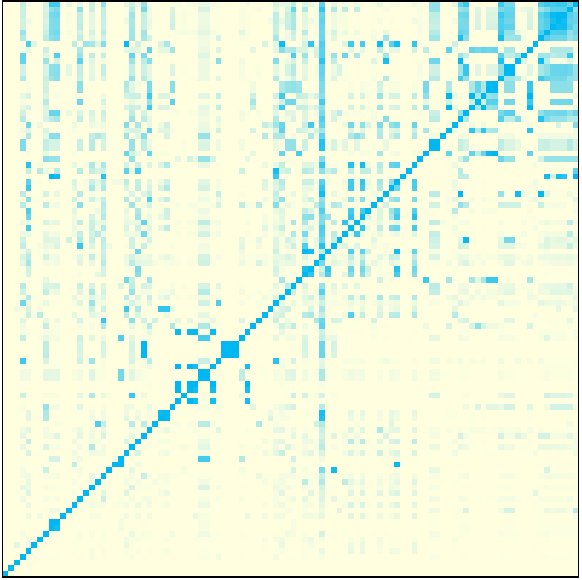}

    \end{subfigure}\hspace*{\fill}
    \begin{subfigure}{0.145\textwidth}

        \includegraphics[width=\textwidth]{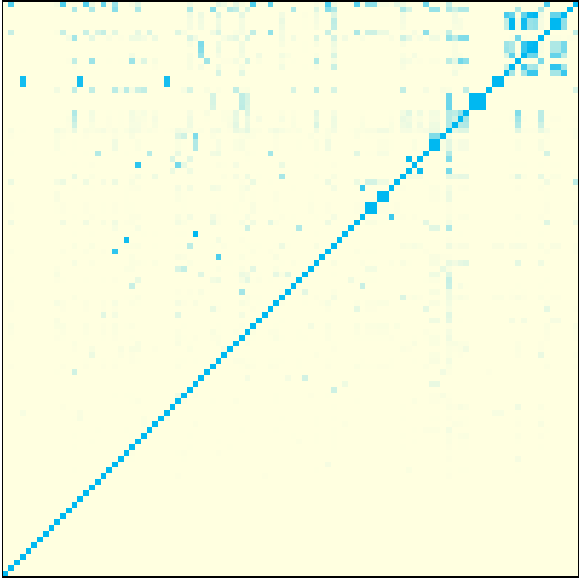}

    \end{subfigure}\hspace*{\fill}
    \begin{subfigure}{0.145\textwidth}

        \includegraphics[width=\textwidth]{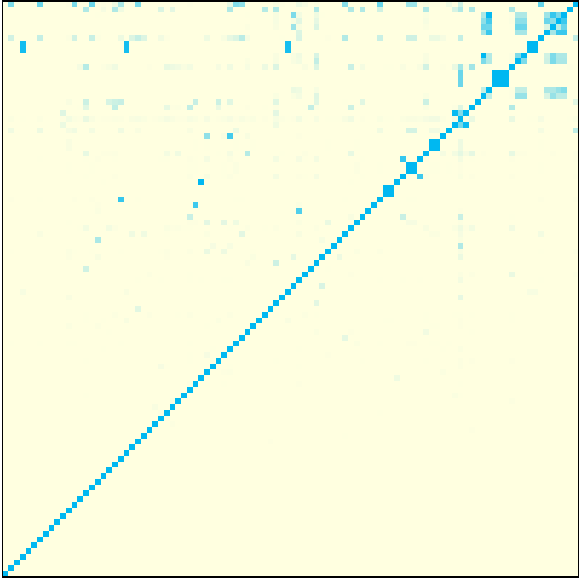}
        \end{subfigure}
    
    \medskip

    \hspace{1.2cm}
    \begin{subfigure}{0.085\textwidth}

        \hspace{-0.3cm}
        \includegraphics[width=\textwidth]{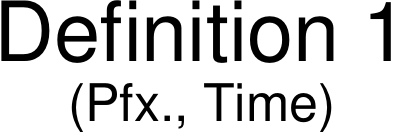}

    \end{subfigure}\hspace*{0.7cm}
    \begin{subfigure}{0.1\textwidth}

        \centering
        \includegraphics[width=\textwidth]{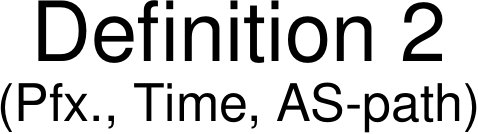}

    \end{subfigure}\hspace*{0.5cm}
    \begin{subfigure}{0.14\textwidth}

        \centering
        \includegraphics[width=\textwidth]{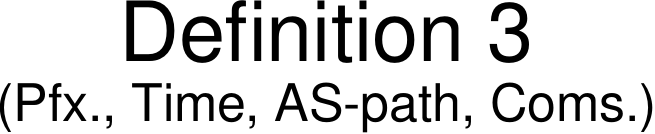}

    \end{subfigure}

    \caption{Redundancy among a subset of 100 existing VPs selected using two different
    techniques for three increasingly stricter redundancy definitions.
    Randomly selecting VPs (top row) returns significantly more pairs of redundant VPs.
    }
    \label{fig:redundancy}
\end{figure}

\myitem{Redundant pairs of VPs exist.}
\fref{fig:redundancy} (top row) shows the level of redundancy for the three definitions and
between 100 VPs randomly selected
and computed over the updates observed during two hours on August 1, 2022.
Observe that we performed 30 random selections with different seeds and show the median case (in terms of redundant pairs of VPs).
One cell in the matrix indicates the redundancy of the VP on the ordinate with the VP on the abscissa.
We define the redundancy between VP $1$ and VP $2$ as the proportion of updates observed 
by VP $1$ that are redundant with at least one update observed by VP $2$.
For better visibility, we show the most redundant VPs at the top of the figures.

Redundant pairs of VPs exist regardless 
of the redundancy definition used. Logically, the stricter the definition, the fewer redundant pairs of VPs.
\fref{fig:redundancy} (left) shows that the VPs can be highly redundant
when they are selected randomly. 
For instance, with the loose Def.
\ref{def:red1}, we observe that 74 among the 100 randomly selected VPs have >50\% of their updates
that are redundant with the ones observed by two other VPs or more
(23 for Def. \ref{def:red2} and 16 for Def. \ref{def:red3}). 
We observe a similar redundancy level when considering only full feeders.

%% file: challenge.tex
\section{Main challenge: prevent overfitting}
\label{sec:challenges}

Our design objective is a general framework that can accommodate
different definitions of redundancy in selecting
the set of least redundant VPs.
 However, optimizing selection
for one objective is likely to overfit, leading to
poor performance for other objectives. 
Thus, while the three definitions in \S\ref{sec:opportunity} enable illustrating the redundancy
across current VPs, none of them are used in the design of \name. These definitions
are too naive to accurately quantify redundancies between the VPs.

We explore this risk of overfitting to a particular objective 
using a VPs selection strategy optimized for one objective:
minimizing redundancy. 
This selection strategy, which we name \textit{greedy specific},
iteratively selects (in a \textit{greedy} fashion) the
VP that minimizes the proportion of redundant updates
across all the updates collected by the selected VPs.
We implement three versions of it, 
one for each redundancy definition used in \S\ref{sec:opportunity}.
Thus, \textit{greedy specific} approximates an optimal 
VP selection when the goal is to minimize redundancy between VPs
according to a specific definition of redundancy.

\myitem{Greedy specific limits redundancy.}
We select 100 VPs using \textit{greedy specific}.
Logically, the selected VPs are
less redundant (see \fref{fig:redundancy}, bottom row) compared to the 100 VPs randomly selected.
With the loose Def. \ref{def:red1}, only 30
VPs have >50\% of their updates redundant with ones observed by two other VPs or more.
This number drops to 9 with Def. \ref{def:red2} and 5 with Def. \ref{def:red3}.
This result highlights that while VPs can be highly redundant, nonredundant pairs of VPs also exist.

\myitem{Greedy specific overfits.}
\textit{Greedy specific} overfits because it optimizes one particular objective.
Thus, it works well for this objective but not for the others. 
We confirm this overfitting effect in \S\ref{sec:eval} where we benchmark \textit{greedy specific} against 
\name on various objectives and show that it performs poorly on objectives that it does not optimize.  
Consequently, one would need to design a \textit{greedy specific} VPs selection for every
possible definition of data redundancy---which is unpractical given that there is an infinite number of definitions.

%% file: overview-kc.tex
\section{Methodology Overview}
\label{sec:solution_overview}

\name samples BGP updates from \ris and \rv at the VP granularity. 
Our method has four steps that we overview below.

\myitem{Step 1 (\S\ref{sec:design_events}): Select a large, unbiased set of BGP events
that we use to gauge pairwise redundancy between VPs.}
\name evaluates the redundancy between two VPs based on 
a carefully selected set of {\em non-global} BGP events (\ie AS path changes).
Global events are typically seen by all VPs and have 
the same impact on every VP view, rendering them
less discriminating for this purpose.  We stratify our selection
of sampled events across space and time to avoid bias. 

\myitem{Step 2 (\S\ref{sec:design_quantification}): Characterize how VPs experience the selected events.}
For every BGP event, \name quantifies topological 
features~\cite{nature_2019} of the ASes involved as observed by
each VP.  These features
embed information about the four attributes of a BGP
update: time, prefix, AS path, and communities. 

\myitem{Step 3 (\S\ref{sec:design_similarity}): Compute pairwise redundancy between VPs.}
\name computes the pairwise Euclidean distance in a $n$-dimensional space, 
where $n$ is the number of topological features times the number of events.
VP pairs with similar feature values for many events are close 
in this space and thus likely redundant.  
\name then computes the average Euclidean distances between 
each pair of VPs computed over different and nonoverlapping 
time periods.

\myitem{Step 4 (\S\ref{sec:design_generation}): Sort and select the least redundant VPs.}
\name relies on a greedy algorithm that considers 
both data \textit{redundancy} and 
its \textit{volume} to build a set of the most valuable VPs.
\name first adds the VP with the lowest
average Euclidean distance to all other VPs, 
and then greedily adds the VP that balances minimal redundancy 
with already selected VPs and minimal additional data volume 
that the VP brings.

%% file: design.tex
\section{Methodology Details}
\label{sec:design}

In the following, we consider the set of VPs
$V$ that includes all VPs from \ris and \rv.
We compute the RIB of VP $v$ at time $t$
using its last RIB dump before $t$ and subsequent updates until $t$.
We use this RIB to construct and maintain 
the undirected weighted graph $G_v(t) = (N_v(t), E_v(t))$
from the AS paths of the best routes observed by $v$ at time $t$, with 
$N_v(t)$ the set of nodes and $E_v(t)\in N_v(t)*N_v(t)$ the set of AS links.
The edges are undirected because
two identical paths in opposite directions should not appear as nonredundant.
Each edge in $E_v(t)$ has a weight in $\mathbb{Z^+}$
which is the number of routes in the RIB that includes this edge 
in their AS path.

\subsection{Select BGP events to assess redundancy}
\label{sec:design_events}

\begin{table}[t]
    \centering

    \resizebox{8.7cm}{!}{
        \def\arraystretch{1.5}
    \begin{tabular}{l|l|l|l|l}
    ID & Name & \# of ASes & Avg.degree & Description \\ \hline 
    1 & Stub & 63310 & 3 & \makecell[l]{ASes without customer} \\
    2 & Transit-1 & 10845 & 27 & \makecell[l]{Transit ASes with a customer \\cone size lower than the average} \\
    3 & Transit-2 & 704 & 267 & \makecell[l]{Transit ASes $\notin$ Transit-1} \\
    4 & HyperGiant & 15 & 1078 & \makecell[l]{Top 15 as defined in~\cite{hypergiant}} \\
    5 & Tier1 & 19 & 1817 & \makecell[l]{Tier1 in the CAIDA dataset~\cite{CAIDA_relationships}} \\
    \end{tabular}
    }

    \vspace{0.2cm}
    \caption{\name balances selected events across 5 AS types.}
    \label{tab:category}

\end{table}

\myitem{\name uses local and partially visible new-AS-link events.}
\name focuses on BGP events that trigger
a new AS link to appear in the path to reach prefix $p$ from different VPs.
A new-AS-link event is a candidate event in $\mathcal{C}$
if at least two and fewer than half of the VPs begin to use
the same new AS link to reach the same prefix within a 10-minute window
(to accommodate typical BGP convergence and path exploration delays~\cite{transientConv,pathexplo2006}).
Since the aim of \name is to find data unique to individual VPs,
we exclude global events (\ie seen by most VPs) to focus on local events.

\myitem{\name avoids biases across time and location.}
From candidate set $\mathcal{C}$, \name builds the final set of events
$\mathcal{E}$ by selecting 15 events
in \SI{500} different and nonoverlapping 10-minute time periods. 
Adding more periods does not
affect significantly the results.
\name samples time periods randomly within a one-month 
timeframe to avoid mis-inferring one larger event 
(\eg a route leak that continuously generates new links for multiple hours)
as several smaller AS-link-level events. 
Inspired by previous approaches to mitigate the risk of 
over-sampling core or stub (edge) ASes~\cite{RIS_pavlos,biases},
our approach classifies ASes into five categories
(\tref{tab:category}) and selects an equal number of 
new-AS-link events for every pair of AS categories.
We distinguish two classes of transit providers by 
customer cone size (Transit-1 and -2) since they have
different topological properties. 
If an AS belongs to more than one category, we classify it in the
category with the highest ID.  
ASes classified
in a lower row of \tref{tab:category} have a higher degree, and there
are more low-degree ASes than high-degree ASes.

\begin{figure}[t]
    \centering
    \begin{adjustbox}{minipage=\linewidth,scale=0.95}
      \centering
      \begin{subfigure}[t]{0.9\textwidth}
        \centering
        \includegraphics[width=1\textwidth]{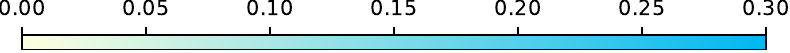}
        \vspace{-0.3cm}

    \end{subfigure}
      \begin{subfigure}[c]{.49\textwidth}
        \centering
  
        \includegraphics[width=0.98\textwidth]{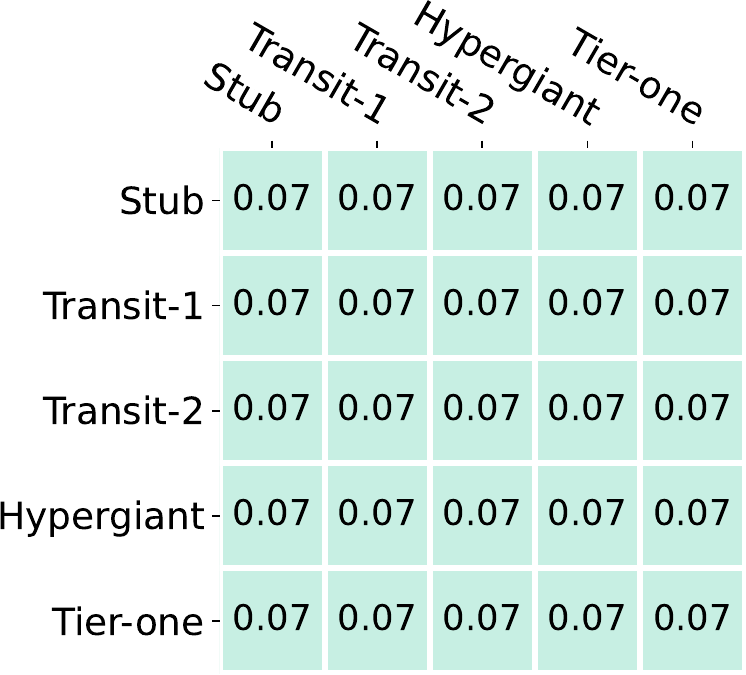}
        \vspace{-1mm}
        \captionsetup{width=1\textwidth}
        \caption{Balanced selection.}
        \label{fig:actual_sampling}
      \end{subfigure}
      \hfill
      \begin{subfigure}[c]{.49\textwidth}
        \centering
  
        \includegraphics[width=0.98\textwidth]{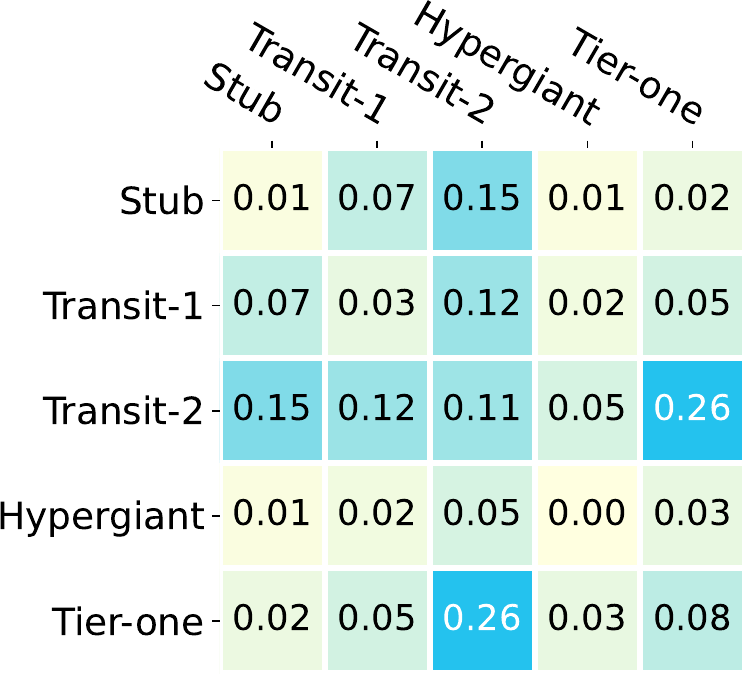}
        \vspace{-1mm}
        \captionsetup{width=1\textwidth}
        \caption{Random selection.}
        \label{fig:possible_sampling}
  
      \end{subfigure}

    \end{adjustbox}
        \captionsetup{width=1\linewidth}
      \caption{\name selects the new-AS-link events using a balanced selection scheme that reduces bias 
      (\fref{fig:actual_sampling} vs.~\fref{fig:possible_sampling}).  The x- and y-axis are the 
      five categories of ASes (see \tref{tab:category}).}
      \label{fig:sampling}
  \end{figure}

\fref{fig:sampling} shows the proportion of selected events
for each of the 15 pairs of AS category (the matrixes 
are symmetric) and for 7500 events 
selected in January 2023 using two schemes: balanced and random.
The random selection (\fref{fig:possible_sampling})
selects many more events involving Transit-2 ASes (69\%)
than hypergiants (11\%), while our 
balanced selection scheme mitigates biases by selecting the same 
number of links in every category (\fref{fig:actual_sampling}).
For each time period, \name selects one event in each of the
15 pairs of AS, yielding $15*500=7500$ events 
($|\mathcal{E}| = 7500$) for use in the next step.

\subsection{Quantifying the observation of the VPs}
\label{sec:design_quantification}

\myitem{\name considers the four main BGP attributes.}
\name computes topological features on the graphs $G_v(t)$ 
for all VPs. The combination of these topological features
prevents overfitting as the graphs on which they are computed 
embed information about the four main BGP attributes (\S\ref{sec:background}).
More concretely, the
graphs $G_v(t)$ embed information about (i) the \textit{time} as the graph
is updated over time, (ii) the \textit{AS path} as it is used to build
the AS graph, (iii) the \textit{prefixes} as they are used to weight
every edge on the graph, and
(iv) the \textit{community values} as they are strongly correlated
with the AS path. We confirm this correlation by downloading the first RIBs of Jan. 2023
for all VPs and analyzing the correlation between the AS path and
the set of BGP communities. We find that two identical AS paths
share the exact same set of BGP communities in 93\% of the cases.
We thus do not embed more information about BGP communities because 
many of them encode local traffic engineering
decisions~\cite{donnet2008} that could lead to \name overfitting.  
We validate this design choice in \S\ref{sec:eval_performance}.

\input{table-features}

\myitem{\name uses 15 diverse topological features
(\tref{tab:descfeatures}).}
\name computes topological features (extracted from literature \cite{topoFeatBook})
that are either node-based or link-based. Node-based features are computed for the two
ends of a new AS link, while link-based are computed for the new AS link.
\name uses six node-based features that we classify
into three categories. The first one quantifies how central and
connected a node is in the graph; the second quantifies how
connected are the neighboring nodes; and the third quantifies
the topological patterns (e.g., triangles) that include the node.
We classify the three pair-based features into a single category
that measures how close two nodes are based on
their neighboring nodes. 
Five features rely on edge weights.
We omit other topological features as they are redundant with the
selected ones. 

\myitem{\name computes the value of the features for each VP 
and selected event.} Consider the event $e \in \mathcal{E}$ that is
the appearance of the AS link $(e_{AS1}, e_{AS2})$ at time $e_t$, 
and the VP $v \in V$.
Computation of the feature values
depends on the feature type.
We denote $F_n$ (resp. $F_p$) the set of node-based (resp. pair-based) features
and show how \name computes the value of these
two types of features for event $e$ and VP $v$.

\noindent
\underline{\textit{Node-based features:}} Consider feature $f_i \in F_n$ and
$f_i(x, G_v(t))$ its value for node $x$ on the graph $G_v(t)$, with $i$
the feature index in \tref{tab:descfeatures}. 
\name computes the following 12-dimensional
feature vector.
\begin{align*}
    T_{node\_based}(v, e) = [f_0(e_{AS1}, G_v(e_t)), f_0(e_{AS2}, G_v(e_t)), \\ 
    \dots, f_5(e_{AS1}, G_v(e_t)), f_5(e_{AS2}, G_v(e_t))]
\end{align*}%
\noindent
\underline{\textit{Pair-based features:}} Consider feature $f_i \in F_p$ and
$f_i(x_1, x_2, G_v(t))$ its value for the node pair $(x_1, x_2)$ on the graph $G_v(t)$, with $i$
the feature index in \tref{tab:descfeatures}. 
\name computes the following 3-dimensional
feature vector.
\begin{align*}
    T_{pair\_based}(v, e) = [f_6(e_{AS1}, e_{AS2}, G_v(e_t)), \\ 
    \dots, f_{8}(e_{AS1}, e_{AS2}, G_v(e_t))]
\end{align*}%
\noindent
The final feature vector used by \name is $T(v, e)$, an 15-dimensional vector 
that is the concatenation (denoted $\oplus$) of the node- and pair-based features.
\begin{align*}
    T(v, e) = T_{node\_based}(v, e) \oplus T_{pair\_based}(v, e)
\end{align*}%

\subsection{Redundancy scoring}
\label{sec:design_similarity}

\name computes pairwise redundancy between VPs in the following four steps. 

\myitem{Step 1: Concatenate the feature vectors.}
\name first concatenates the computed topological feature vectors (15 features) for all the events selected
in the same time period (15 events).
We denote $\mathcal{E}_p$ the events selected in the \textit{p}-th time period ($|\mathcal{E}_p|=15$),
with $0 \leq p < 500$, and denote $e_{p, i} \in \mathcal{E}_p$ the \textit{i}-th selected event in the \textit{p}-th time period. 
$F(v, p)$ is the concatenated feature vector 
for VP $v$ and the events $\mathcal{E}_p$, which has $15*15=225$ 
dimensions and which \name calculates as:
\begin{align*}
    F(v, p) = T(v, e_{p,0}) \oplus T(v, e_{p,1}) \oplus \dots \oplus T(v, e_{p, 14})
\end{align*}%
\myitem{Step 2: Normalize concatenated feature vectors.}
\name normalizes the data for each time period using the feature matrix $\mathcal{M}(p)$ that includes 
the concatenated feature vectors for all VPs (rows) and events (columns) 
in period $p$.
\begin{equation*}
    \renewcommand{\arraystretch}{1}
        \label{eq:features}
    \mathcal{M}(p) =
    \begin{bmatrix}
        F(v_{0}, p) \\
        \dots    \\
        F(v_{|V|}, p)

    \end{bmatrix}
\end{equation*}
\noindent
\name normalizes (operation $\bigtriangledown$) the matrix $\mathcal{M}(p)$ column-wise using
a standard scaler that transforms every column such
that its average is zero and its standard deviation is one.

\myitem{Step 3: Compute Euclidean distance between VPs.}
\name uses the normalized matrix $\bigtriangledown (\mathcal{M}(p))$
to compute the Euclidean distance between 
every pair of VPs and for all events in the time period $p$ (operation $\diamond$).
We denote $\bigtriangledown(\mathcal{M}(p))_x$ the \textit{x}-th row in the matrix $\bigtriangledown(\mathcal{M}(p))$
and $\bigtriangledown(\mathcal{M}(p))_{x,i}$ its value at index $i$ (\ie the \textit{i}-th column).
We define the Euclidean distance between the \textit{n}-th VP $v_n$ and the \textit{m}-th VP $v_m$ over
the selected events in the time period $p$ as follows.
\begin{align*}
    \diamond(v_n, v_m, p) = \sum_{i=0}^{225} (\bigtriangledown(\mathcal{M}(p))_{n, i} - \bigtriangledown(\mathcal{M}(p))_{m, i})^{2}
\end{align*}%
\myitem{Step 4: Compute the average distance over all time periods.}
The redundancy score $\mathcal{R}(v_n, v_m)$ between two VPs $v_n$ and $v_m$
relates to the normalized average Euclidean 
distance between them over the 500 time periods, computed as:
\begin{align*}
    \mathcal{R}(v_n, v_m) = 1 - \coprod((\sum^{500}_{p=0}{\diamond(v_n, v_m, p)})*\frac{1}{500})
\end{align*}%
\noindent
The operator $\coprod$ applies a min-max scaler so that scores are between 0 and 1,
with 1 meaning the most redundant pair of VPs and 0 the less redundant pair of VPs.  
\name thus computes and returns a redundancy score for every pair of VPs.

\subsection{Generating a set of VPs}
\label{sec:design_generation}

We now explain how \name generates a set of
VPs $\mathcal{O}$ that minimizes the proportion of redundant information collected.
\name initializes the set $\mathcal{O}$ with the most 
redundant VP, \ie the one with the lowest sum of Euclidean distances to all the other VPs.
This design choice allows the redundant part of the BGP data (\eg c2p links) to be visible by the first selected VP.  
At every following iteration, \name builds a candidate set of VPs $\mathcal{K}$ 
that contains the unselected VPs exhibiting the lowest maximum redundancy score.
The maximum redundancy score $P$
measures the maximum redundancy between a VP $v$
and the set of VPs $\mathcal{O}$ and is defined as follows.   
\begin{equation*}
    P(\mathcal{O}, v) = \max(\mathcal{R}(v, v_i), \forall v_i \in \mathcal{O})
\end{equation*}
\name adds in $\mathcal{K}$ the $\alpha= $ 25\% of the nonselected VPs that
exhibit the lowest maximum redundancy score.

\name then adds in set $\mathcal{O}$ the VP 
that is in the candidate set $\mathcal{K}$ and that collects the lowest volume of data
compared to the other VPs in $\mathcal{K}$.
\name estimates the volume of data collected by the VPs by counting 
the number of updates that they received over 365 one-hour periods, one randomly
selected in each day of the year
to align with the yearly update rate of \name (\S\ref{sec:functions}).
The $\alpha$ parameter allows tuning redundancy and volume knobs:
a low $\alpha$ prioritizes low redundancy 
while a higher $\alpha$ prioritizes low resulting data volume.
We found that $\alpha = $ 25\% performs well in practical
scenarios (we tested a range from 10\% to 50\%).

%% file: table-features.tex
\begin{table}[ht]
  \centering

  \resizebox{8.5cm}{!}{
  \def\arraystretch{1}

  \begin{tabular}{ccccc}
      \toprule
      Type & Categorie & Name & Weighted & Index \\\cmidrule{1-5}\vspace{0.1cm}
      \multirow{8}{*}{{\rotatebox[origin=c]{90}{ \textbf{Node-based}}}} & \multirow{2}{*}{\textbf{Centrality Metrics}}  & Closeness centrality & \darkgreencheck & 0 \\\vspace{0.1cm}
      & & Harmonic centrality & \darkgreencheck & 1 \\ \cmidrule{2-5}\vspace{0.1cm}
      & \multirow{2}{*}{\textbf{Neighborhood Richness}} & Average neighbor degree & \darkgreencheck & 2 \\\vspace{0.1cm}
      & & Eccentricity & \darkgreencheck  & 3 \\\cmidrule{2-5}\vspace{0.1cm}
      & \multirow{2}{*}{\textbf{Topological Pattern}}& Number of Triangles & \darkredcross & 4 \\\vspace{0.1cm}
      & & Clustering & \darkgreencheck & 5 \\\toprule \vspace{0.1cm}
      \multirow{3}{*}{{\rotatebox[origin=c]{90}{ \textbf{Pair-based}}}} & \multirow{3}{*}{\textbf{Closeness Metrics}} & Jaccard & \darkredcross & 6 \\\vspace{0.1cm}
      &  & Adamic Adar & \darkredcross & 7 \\\vspace{0.2cm}
      & & Preferential attachment & \darkredcross & 8 \\\toprule

  \end{tabular}
  }

  \caption{Node-based and pair-based features used by \name.}
  \label{tab:descfeatures}
\end{table}

%% file: software.tex
\section{System functionalities}
\label{sec:functions}

\name runs on a commodity server.
Upon launch, it collects BGP routes from \ris and \rv using BGPStream~\cite{bgpstream}
and computes the redundancy between every pair 
of VPs at a yearly granularity, which is enough given that redundancies between VPs 
remain stable over time (see \S\ref{sec:eval_sound}).
\name then takes as input a \textit{year}
and a \textit{volume of data} and returns a set of  VPs that generates
a volume of data lower than the volume specified as input.
\name returns the redundancy scores
calculated for every pair of VPs.
Thus, users have the option to compute their own set
of complementary VPs based on these redundancy scores and some additional constraints that they might have.
This is useful when users want to include (or exclude) some VPs (regardless of how redundant they are), 
which will result in another set of VPs rather than the default set
provided by \name.
For instance, when trying to detect new peering, a user may
want to take some VPs at an IXP in addition to some VPs selected
by \name.

\name runs at \website,~allowing users 
to get a list of VPs or the redundancy scores without computational expenses.
We implemented three versions of \name, one for IPv4 routes (\namevfour), one
for IPv6 routes (\namevsix) and one that considers both IPv4 and IPv6 routes (\namevboth).
The three versions use the same methodology (described in \S\ref{sec:design})
to compute redundancy and generate a set of VPs.

%% file: evaluation.tex
\section{Evaluation}
\label{sec:eval}

We show that \name improves the trade-off between the volume of data
collected and the routing information inferred compared to current VPs selection strategies 
in five use cases for which we have ground truth (\S\ref{sec:eval_performance}). 
We then show that \name would improve coverage and accuracy of
previous studies for which ground truth is unknown (\S\ref{sec:eval_impact}).
Finally, we show that the key design choices of \name are sound (\S\ref{sec:eval_sound}).

\begin{table*}[!t]
	\centering
	\captionsetup{width=15cm}

	\resizebox{15.5cm}{!}{

	\begin{tabular}{cc|ccc|ccccc|ccc}
		\Xhline{1.5pt}
		 \multirow{2}{*}{Use case} & \multirow{2}{*}{Objective} & \multicolumn{3}{c|}{Naives baselines} & \multicolumn{5}{c|}{Greedy specifics use cases (\S\ref{sec:eval_performance})} & \multicolumn{3}{c}{Greedy specifics Def. (\S\ref{sec:opportunity})}\\\cline{3-6}\cline{7-13}
		 &  & Random & AS-distance & \bb~\cite{RIS_pavlos} & \First & \Second & \Third & \Fourth & \Fifth & Def. 1 & \hspace{0.3cm}Def. 2 & Def. 3\\\Xhline{1.5pt}\vspace{0.0cm}
		\multirow{4}{*}{\vspace{0.4cm}\fboxrule=0pt\fbox{\parbox[c]{2.1cm}{\small \centering \textbf{Transient path detection\\(\First)}}}} & 50 \% & \ApplyGradient{1.55} & \ApplyGradient{1.76} & \ApplyGradient{1.82} & \ApplyGradient{0.70} & \ApplyGradient{2.99} & \ApplyGradient{3.29} & \ApplyGradient{3.82} & \ApplyGradient{2.89} & \ApplyGradient{1.96} & \ApplyGradient{2.12} & \ApplyGradient{1.69}\\\cline{2-13}\vspace{0.0cm}
		 & 70 \% & \ApplyGradient{1.38} & \ApplyGradient{1.62} & \ApplyGradient{1.53} & \ApplyGradient{0.76} & \ApplyGradient{3.24} & \ApplyGradient{3.51} & \ApplyGradient{3.42} & \ApplyGradient{3.09} & \ApplyGradient{1.56} & \ApplyGradient{1.56} & \ApplyGradient{1.78}\\\cline{2-13}\vspace{0.0cm}
		 & 90 \% & \ApplyGradient{1.13} & \ApplyGradient{1.17} & \ApplyGradient{1.21} & \ApplyGradient{0.75} & \ApplyGradient{1.66} & \ApplyGradient{1.67} & \ApplyGradient{1.66} & \ApplyGradient{1.66} & \ApplyGradient{1.33} & \ApplyGradient{1.15} & \ApplyGradient{1.59}\\\Xhline{1.5pt}\vspace{0.0cm}
		\multirow{4}{*}{\vspace{0.4cm}\fboxrule=0pt\fbox{\parbox[c]{2.1cm}{\small \centering \textbf{MOAS detection\\(\Second)}}}} & 50 \% & \ApplyGradient{2.35} & \ApplyGradient{3.38} & \ApplyGradient{3.41} & \ApplyGradient{2.31} & \ApplyGradient{0.98} & \ApplyGradient{1.80} & \ApplyGradient{2.83} & \ApplyGradient{1.53} & \ApplyGradient{3.39} & \ApplyGradient{2.85} & \ApplyGradient{3.98}\\\cline{2-13}\vspace{0.0cm}
		 & 70 \% & \ApplyGradient{2.18} & \ApplyGradient{3.44} & \ApplyGradient{3.38} & \ApplyGradient{2.56} & \ApplyGradient{0.85} & \ApplyGradient{1.79} & \ApplyGradient{2.30} & \ApplyGradient{1.83} & \ApplyGradient{3.02} & \ApplyGradient{2.66} & \ApplyGradient{3.67}\\\cline{2-13}\vspace{0.0cm}
		 & 90 \% & \ApplyGradient{1.98} & \ApplyGradient{2.69} & \ApplyGradient{3.06} & \ApplyGradient{2.37} & \ApplyGradient{1.04} & \ApplyGradient{2.31} & \ApplyGradient{2.82} & \ApplyGradient{2.56} & \ApplyGradient{2.46} & \ApplyGradient{2.19} & \ApplyGradient{3.31}\\\Xhline{1.5pt}\vspace{0.0cm}
		\multirow{4}{*}{\vspace{0.4cm}\fboxrule=0pt\fbox{\parbox[c]{2.1cm}{\small \centering \textbf{AS topology mapping\\(\Third)}}}} & 50 \% & \ApplyGradient{2.59} & \ApplyGradient{2.97} & \ApplyGradient{2.43} & \ApplyGradient{1.58} & \ApplyGradient{1.29} & \ApplyGradient{0.71} & \ApplyGradient{1.53} & \ApplyGradient{1.94} & \ApplyGradient{2.27} & \ApplyGradient{2.18} & \ApplyGradient{3.35}\\\cline{2-13}\vspace{0.0cm}
		 & 70 \% & \ApplyGradient{2.06} & \ApplyGradient{2.29} & \ApplyGradient{2.13} & \ApplyGradient{1.33} & \ApplyGradient{1.22} & \ApplyGradient{0.64} & \ApplyGradient{1.29} & \ApplyGradient{1.37} & \ApplyGradient{2.14} & \ApplyGradient{1.64} & \ApplyGradient{2.28}\\\cline{2-13}\vspace{0.0cm}
		 & 90 \% & \ApplyGradient{1.72} & \ApplyGradient{1.88} & \ApplyGradient{1.80} & \ApplyGradient{1.30} & \ApplyGradient{1.18} & \ApplyGradient{0.77} & \ApplyGradient{1.23} & \ApplyGradient{1.27} & \ApplyGradient{1.73} & \ApplyGradient{1.64} & \ApplyGradient{1.80}\\\Xhline{1.5pt}\vspace{0.0cm}
		\multirow{4}{*}{\vspace{0.4cm}\fboxrule=0pt\fbox{\parbox[c]{2.1cm}{\small \centering \textbf{Traffic engineering detection\\(\Fourth)}}}} & 50 \% & \ApplyGradient{4.59} & \ApplyGradient{4.74} & \ApplyGradient{4.82} & \ApplyGradient{3.76} & \ApplyGradient{2.67} & \ApplyGradient{2.34} & \ApplyGradient{0.47} & \ApplyGradient{3.33} & \ApplyGradient{3.95} & \ApplyGradient{3.34} & \ApplyGradient{4.76}\\\cline{2-13}\vspace{0.0cm}
		 & 70 \% & \ApplyGradient{2.71} & \ApplyGradient{3.37} & \ApplyGradient{3.52} & \ApplyGradient{3.04} & \ApplyGradient{1.86} & \ApplyGradient{2.89} & \ApplyGradient{0.41} & \ApplyGradient{1.85} & \ApplyGradient{4.34} & \ApplyGradient{2.02} & \ApplyGradient{3.51}\\\cline{2-13}\vspace{0.0cm}
		 & 90 \% & \ApplyGradient{1.55} & \ApplyGradient{1.70} & \ApplyGradient{1.95} & \ApplyGradient{1.61} & \ApplyGradient{1.54} & \ApplyGradient{1.52} & \ApplyGradient{0.32} & \ApplyGradient{1.46} & \ApplyGradient{1.88} & \ApplyGradient{1.33} & \ApplyGradient{1.89}\\\Xhline{1.5pt}\vspace{0.0cm}
		\multirow{4}{*}{\vspace{0.4cm}\fboxrule=0pt\fbox{\parbox[c]{2.1cm}{\small \centering \textbf{Unnecessary updates detection\\(\Fifth)}}}} & 50 \% & \ApplyGradient{1.72} & \ApplyGradient{2.89} & \ApplyGradient{2.41} & \ApplyGradient{2.19} & \ApplyGradient{2.10} & \ApplyGradient{2.63} & \ApplyGradient{2.20} & \ApplyGradient{0.38} & \ApplyGradient{2.43} & \ApplyGradient{2.92} & \ApplyGradient{2.59}\\\cline{2-13}\vspace{0.0cm}
		 & 70 \% & \ApplyGradient{1.30} & \ApplyGradient{2.04} & \ApplyGradient{1.91} & \ApplyGradient{1.43} & \ApplyGradient{1.53} & \ApplyGradient{1.50} & \ApplyGradient{1.90} & \ApplyGradient{0.39} & \ApplyGradient{1.51} & \ApplyGradient{1.63} & \ApplyGradient{2.02}\\\cline{2-13}\vspace{0.0cm}
		 & 90 \% & \ApplyGradient{1.01} & \ApplyGradient{1.36} & \ApplyGradient{1.39} & \ApplyGradient{1.17} & \ApplyGradient{1.18} & \ApplyGradient{1.14} & \ApplyGradient{1.16} & \ApplyGradient{0.50} & \ApplyGradient{1.09} & \ApplyGradient{1.38} & \ApplyGradient{1.35}\\\Xhline{1.5pt}\vspace{0.0cm}
	\end{tabular}
	}
	\caption{Data reduction factors with \namevfour compared to several baselines for five use cases.
	\name outperforms every baseline for all five use cases. Unlike \textit{greedy specifics}, \name greatly avoids overfitting.}
	\label{tab:benchmark_v4}
\end{table*}

\subsection{Benchmarking \name}
\label{sec:eval_performance}

We benchmark \name against three baselines per use case.

\myitem{Use cases.}
We evaluate \name on five different use cases that we carefully picked 
such that each BGP attribute is useful for at least one of them.
For instance, the \textit{time} is useful to detect transient
events (use case \First); the \textit{prefix} is useful to
detect Multiple Origin ASes (MOAS) prefixes (use case \Second);
the \textit{AS path} is useful to map the Internet topology
(use case \Third); and the \textit{community values} are useful
to detect traffic engineering (use case \Fourth) and unnecessary
updates (use case \Fifth).  Our goal is to demonstrate that
\name does not overfit on some particular use cases or BGP attributes.
For each use case, we process the updates collected during 
100 one-hour periods (randomly selected in May 2023) and benchmark 
\name on a set of events found. We thus have ground truth. 
We briefly describe below each use case along with our experimental settings.

\noindent
\underline{\First \textit{Transient paths detection.}} Transient paths are 
BGP routes visible for less than five minutes,
a typical BGP convergence delay~\cite{transientConv}, and which
can be attributed to \eg path exploration~\cite{pathexplo2006}.
We focus on 200 randomly selected transient path events for every one-hour period, 
making a total of $100*200=20000$ events used. \\
\underline{\Second \textit{MOAS prefixes detection.}}
MOAS prefixes are announced by multiple distinct ASes~\cite{artemis},
which can be caused by legitimate \cite{moas_legit}
or malicious \cite{youtube_hijack,russian_moas,china_moas} actions. 
We focus on 200 MOAS randomly selected events for every one-hour period,
making a total of $100*200=20000$ MOAS events used. \\
\underline{\Third \textit{AS topology mapping.}} 
This is useful for \eg inferring BGP
policies~\cite{Luckie2013} or AS paths~\cite{ASpathInf}.
For each VP, we process the first RIB dump of May 2023 as well as
the updates collected during the 100 one-hour periods and 
focus on all distinct AS links observed.\\
\underline{\Fourth \textit{Traffic engineering detection.}}
We focus on action communities \ie those associated with traffic engineering actions~\cite{krenc2023}. 
For every one-hour time period, we focus on 80 updates for which a path change
coincides with the appearance of an action community, 
making a total of $100*80=8000$ path changes used.\\
\underline{\Fifth \textit{Unnecessary Updates detection.}}
An unnecessary update is a BGP update that only signals
a change in the community values but not in the AS path~\cite{unnecessary_updates}.
We consider 200 unnecessary updates randomly picked within each one-hour period, 
making a total of $100*200=20000$ events used.

\myitem{Baselines.}
We benchmarked \name against three naive baselines
commonly used in practice (\S\ref{sec:problem2}):
\first \textit{random} selection of VPs, which results in a skewed set of VPs
as they exhibit biases~\cite{RIS_pavlos}; \second \textit{AS-distance}, \ie
select the first VP randomly and the following ones to maximize 
the AS-level distance between selected VPs; and \third \textit{unbiased}, \ie
start with all VPs and iteratively remove the one that most increases 
the bias on the set of remaining VPs. 
We measure the bias using the definition in~\cite{RIS_pavlos}.

We compare \name against the three \textit{greedy specific} VPs selection strategies optimized for 
Def. \ref{def:red1}, \ref{def:red2}, and \ref{def:red3} (\S\ref{sec:opportunity}).
Additionally, we compare \name against five other \textit{greedy specifics},
one optimized for each of the five use cases described above.
Unlike the \textit{greedy specifics} described in \S\ref{sec:challenges},
these five \textit{greedy specifics} optimize the trade-off between
the volume of the data and
its capacity to achieve a particular objective.
For instance, when the objective is to map the AS topology (use case \Third),
\textit{greedy specific}
iteratively selects the VP that best improves the trade-off between
the number of discovered AS links and the volume of processed data.

\myitem{Reduction factor definition.}
We define the \spfa to capture 
how much \name reduces the number of BGP updates required
to fulfill a particular objective.
More precisely, assume an objective $O$ and a baseline $B$. 
We iteratively build a set of VPs using baseline $B$. At every iteration,
we download all the updates that the newly selected VP observes during 100 one-hour
periods randomly selected in May 2023. We stop iterating when all updates 
collected by the selected VPs enable the data to meet $O$. Similarly, we build another
set of VPs using \name and stop selecting new VPs (see \S\ref{sec:design_generation})
when the selected ones meet $O$.  
The \spfa is the ratio between the 
number of updates processed with $B$ and with \name.
More formally, the \spfa is $\frac{|U_B^O|}{|U_{MVP}^O|}$ with $|U_B^O|$ and
$|U_{MVP}^O|$ the number of updates processed to fulfill objective $O$
with baseline $B$ and \name respectively.
A \spfa $=2$ means that we can fulfill objective $O$ with
half as many updates when using \name compared to when using baseline $B$. 
More generally, a \spfa $>$ 1 means that
we can fulfill the same objective
with less data when using \name compared to when using $B$.

\myitem{Benchmark results.}
\tref{tab:benchmark_v4} summarizes our results. 
For each use case, we focus 
on three objectives: mapping X\% of the AS topology (use case \Third) or 
detecting X\% of the events (use case 
\First, \Second, \Fourth, and \Fifth), with
X equal to 50, 70, or 90. Here, we focus on the performance of \namevfour.
\namevsix and \namevboth yield comparable performance (see \S\ref{app:eval}).

	\noindent
\underline{\textit{Takeaway \#1:}}
\name outperforms every naive baseline for every use case, \ie the \spfa is always above one.
For instance, we detect 90\% of the MOAS events 
with 3.06$\times$ less data (the \spfa is 3.06)
when using \name compared to selecting the VPs
using the \textit{unbiased} baseline. This means that \name only 
needs 32\% of the updates required by the \textit{unbiased} baseline to fulfill the objective.
Comparably to what we observe in our mini-Internet simulations
(\S\ref{sec:problems}), the \textit{random} baseline performs
better than \textit{AS-distance}.

\noindent
\underline{\textit{Takeaway \#2:}}
We can see that \name generalizes whereas greedy specific overfits.
In fact, for a particular use case, \name is less performant 
than the \textit{greedy specific} strategy optimized for this use case.
For any other use case, \name performs better
than the \textit{greedy specifics}
not optimized for that use case.   
These results demonstrate that the \textit{greedy specific} strategies 
overfit. They are
also not practical as they need ground truth.

\subsection{Impact on previous works}
\label{sec:eval_impact}

We show that \name would improve
the outcome of three measurement studies
and tools that are fueled by the BGP data from \ris and \rv
(and for which there is no ground truth).

\vspace{-0.15cm}
\subsubsection{Inference of AS properties}\hfill%
\label{sec:eval_as_rel}

We show that \name improves AS relationship inferences
(a popular research problem~\cite{Luckie2013,TopoScope, ComplexRel, Unari})
and AS ranking~\cite{asrank}.

\myitem{\name helps to infer +15\% more AS relationships.}
We replicate the methodology proposed in~\cite{Luckie2013} that relies on public BGP data from \ris and \rv
to infer AS relationships and build the widely-used CAIDA
AS-relationship dataset~\cite{CAIDA_relationships}.
We compute the number of inferred AS relationships for every month in 2023
when using the 648 VPs that CAIDA uses to build its dataset (In January 2023)
and when using VPs selected by \name.
We ensure that the VPs selected with \name generate the same volume of data
as the 648 used by CAIDA so that any performance gap can confidently be attributed to \name.
We find that the VPs selected by \name
enable consistent (from Jan. 2023 to Aug. 2023) inference of 
$\approx$90k additional AS relationships
($\approx$+17\%) while
missing only $\approx$11k AS relationships ($\approx$2.2\%) 
present in the original dataset.
Thus, the tradeoff is largely in favor of using \name ($\approx$+15\% overall).

We also replicated the AS relationship validation algorithm used in~\cite{Luckie2013}
(which relies on the IRR and RIR data) and found that the true positive rate
(the metric used in~\cite{Luckie2013}) remains identical (97\%).
Thus, \name significantly improves coverage
without processing more data or losing accuracy.

\myitem{\name prevents flawed inferences in the ASRank dataset.}
We replicate the methodology used by ASRank~\cite{asrank} to
compute the AS Customer Cone Sizes (CCS). 
We find that the CCS changes for 1067 ASes
when using \name and
manually investigated two cases of substantial changes:

\noindent
\underline{\textit{Case I}}\footnote{https://asrank.caida.org/asns?asn=132337\&type=search}: AS132337 has a CCS of 1 in the original 
dataset and a CSS of 18k when using \name, making it 
the 15th AS highest ranked by CCS.  We contacted AS132337
who confirmed that it has 18k customers.
\name correctly ranks AS132337
because it selects the unique VP that sees it as a transit AS.

\noindent
\underline{\textit{Case II}}:\footnote{https://asrank.caida.org/asns?asn=24745\&type=search}
AS24745 is the route server of Balcan-IX and has a CSS of 16 in the original ASrank dataset.
However, we manually checked its participants
and found that the 16 customers are misclassified and actually peer through AS24745. 
With \name, the CSS of AS24745 is 1 and these errors are avoided.

In both cases, \name enables more accurate inferences of CCSs
because it collects more diverse AS paths.
Thus, we can confidently say that \name would prevent many flawed inferences
likely present in the dataset provided by ASRank.

\vspace{-0.15cm}
\subsubsection{Detection of forged-origin hijacks}\hfill%

We show that \name improves forged-origin hijack detection, which is the goal 
of many systems that use BGP routes from \ris
and \rv~\cite{tma19_bgphijack,artemis,grip,pathendvalidation,dfoh}.
Forged-origin hijacks are a type of BGP hijack where the attacker prepends 
the valid origin to the AS path to make the hijacked route appear legitimate.

\myitem{\name improves the accuracy of forged-origin hijack inferences.}
We replicate the algorithm of DFOH~\cite{dfoh} that uses routes collected by 
287 \ris and \rv VPs to infer forged-origin hijacks.
We implement two versions of DFOH, one called DFOH$_{MVP}$ which
uses a set of VPs selected with \name, and another one called DFOH$_R$ that
uses a random set of VPs. In both versions, we ensure that the volume of data collected 
is identical to the one used in~\cite{dfoh}.
As DFOH relies on probabilistic inference, we measure the performance 
of DFOH$_{MVP}$ and DFOH$_R$ in terms of True Positive Rate (TPR) and False Positive Rate (FPR).
We obtain an approximation of ground truth (needed to compute the TPR and FPR)
by implementing a third version of DFOH, called DFOH$_{ALL}$ that uses 
all VPs from \ris and \rv.
Observe that DFOH$_{ALL}$ is an approximation of ground truth
because incorrect inferences are still possible even if all VPs are used.
We restrict our analysis to one month (Jan. 2022) because DFOH$_{ALL}$
is resource-hungry as it uses all VPs.
We find that DFOH$_{MVP}$ uncovers 947 suspicious cases against only 700 for DFOH$_R$.
DFOH$_{MVP}$ outperforms DFOH$_{R}$ for both the TPR and the FPR:
It has a TPR of 85.7\% (against 61.1\% for DFOH$_R$) and a FPR of 14.4\% (against 60.1\% for DFOH$_R$)---a $\approx$4$\times$ better precision.  

\myitem{DFOH$_R$ misses suspicious cases that DFOH$_{MVP}$ does not.}
We manually investigated, using public peering databases (\eg PeeringDB)
some of the suspicious cases inferred by DFOH$_{MVP}$ and not by DFOH$_{R}$.
We find cases that appear particularly suspicious (thus useful for operators)
and describe two of them below (also found by the original DFOH).

\noindent
\underline{\textit{Case I}}\footnote{http://dfoh.uclouvain.be/cases/2022-01-01\_1239\_267548}: 
On Jan. 1, 2022, AS267548, a small Peruvian AS, appears between Sprint,
a Tier1 AS, and AS199524, a large content provider.
However, AS267548 is not supposed to provide transit between these two ASes.

\noindent
\underline{\textit{Case II}}\footnote{http://dfoh.uclouvain.be/cases/2022-01-06\_9269\_268568}: 
On Jan. 6, 2022, AS9269, an ISP based in Hong Kong appears directly connected with AS268568, a Brazilian ISP.
These two ASes do not share any IXP and are not supposed to peer directly.

These two cases show that \name enables the detection of 
additional potential routing attacks versus not using it.

\vspace{-0.15cm}
\subsubsection{Characterizing international routing detours}\hfill%
\label{sec:eval_detours}

\begin{table}[t]
    \centering
    \resizebox{8.5cm}{!}{
    \begin{tabular}{lcccc}
        \cmidrule[1.5pt]{1-5}
        Experiment & Duration & \# of VPs & \# of processed Updates & \# of Detours\\\cmidrule{1-5}
        Original paper & 1 Month & All VPs & $\approx$61B & 174k\\\cmidrule[1.5pt]{1-5}
		\multirow{2}{*}{Random selection} & 2 Months & 624 (median) & $\approx$61B & 165k (median)\\\cmidrule{2-5}
		 & 4 Months & 313 (median) & $\approx$61B & 171k (median)\\\cmidrule[1.5pt]{1-5}
        \multirow{2}{*}{\name selection} & 2 Months & 413 & $\approx$61B & \textbf{250K}\\\cmidrule{2-5}
         & 4 Months & 220 & $\approx$61B & \textbf{263k}\\\cmidrule[1.5pt]{1-5}
    \end{tabular}
    }
    \caption{Using fewer VPs selected by \name
	enables a longer study that detects more detours
	with the same volume of data.}
    \label{tab:mvp_routing_detours}
\end{table}
We focus on a study that uses all VPs to characterize
international routing detours over one month~\cite{forwarding-detour}.
International detours occur when two ASes in the same country are
reachable through an AS in another country, which can lead to
extra forwarding delays.
We show that by using fewer VPs selected by \name, we can lengthen the duration of the study
to find more detours 
without processing more data.

\newpage
\myitem{\name helps to detect +44\% more routing detours.}
We replicate the methodology used in~\cite{forwarding-detour} to detect routing detours
except that \first we use a set of VPs selected using \name
that generates $\alpha\times$ less data  
compared to using them all, with $\alpha=2$ and $\alpha=4$, and
\second we run the analysis over two months when $\alpha=2$
and four months when $\alpha=4$. Thus, the overall volume of data collected remains similar ($\approx$61B RIB entries), regardless of $\alpha$. 
\tref{tab:mvp_routing_detours} shows the number of routing detours detected
in May 2023 (and until June and August 2023 when $\alpha=2$ and 4, respectively).
We detect 250k detours 
over two months ($\alpha=2$) when using 413 VPs selected by \name---a +44\% increase
compared to using all VPs during one month as in~\cite{forwarding-detour}.
When $\alpha=4$, we use 220 VPs selected by \name on four months and find 263k
detours---better than using them all on one month.

We explored the trade-off between the number of VPs 
and the duration of the study using a \textit{random} VPs selection strategy. We detected 
165k detours when using $\approx624$ random VPs
and running the analysis over two months (we tested
the \textit{random} selection with 50 seeds and report the median in \tref{tab:mvp_routing_detours}).
This is fewer than when we replicated the original experiment,
which demonstrates 
that optimized VP selection enables discovering more routing detours.

\myitem{\name enables improved characterization of routing detours.}
We replicate the methodology used in~\cite{forwarding-detour} to
rank countries based on their number of detours,
and ASes based on how often they originate a detoured path.
We find differences when using \name, including two interesting cases: 

\noindent
\underline{\textit{Case I:}} 
Using \name (with $\alpha=2$), 
we discover 33k (+68\%) additional detours traversing the US and 22k (+37\%) traversing Russia 
compared to when using the settings in~\cite{forwarding-detour}.
These additional detours rank the US as the \#1
country with the highest number of routing detours and Russia
as \#2, whereas with the settings in ~\cite{forwarding-detour} Russia is ranked \#1 and the US \#2.

\noindent
\underline{\textit{Case II:}}
Using \name (with $\alpha=2$) enables detecting 720 (+83\%) additional 
routing detours involving AS262503 compared to when
using the settings of~\cite{forwarding-detour}.  This changes rankings:
AS262503 became \#1 vs.~\#7 with the settings in~\cite{forwarding-detour}.

As our rankings are based on the highest number of routing detours 
compared to~\cite{forwarding-detour}, we can confidently say \name improves
the characterization of international routing detours.

%% file: soundness_design.tex
\subsection{Soundness of design choices}
\label{sec:eval_sound}

We show that our three key design choices -- yearly update frequency of 
redundancy scores, balanced sampling, and topological feature selection --
are sound.

\myitem{\name's redundancy scores are sufficiently stable over time that annual recomputation is sufficient}
We ran \name every six months, starting in January 2023 and then going backward 
until January 2018 (\ie a total of ten independent runs).
We limit the scope of this experiment to 100 randomly selected VPs
to limit the computational resources required. 
Logically, we find that the redundancy score differences increase as the time interval between 
two runs of \name increases. However, these differences are low.
The median difference between the scores of two runs of \name separated by one year is only 0.021
(which corresponds to a difference of 9\%), and it increases to 0.171 (\ie a difference of 23\%)
when the two runs are separated by four years.
We thus configure \name to recompute
redundancy scores and update its set of selected VPs on a yearly basis (see \S\ref{sec:functions})---a
good trade-off between computational cost and performance.

\begin{figure}
    \begin{minipage}{0.45\textwidth}
        \begin{minipage}{0.49\textwidth}
            \centering
            \includegraphics[width=1.0\textwidth]{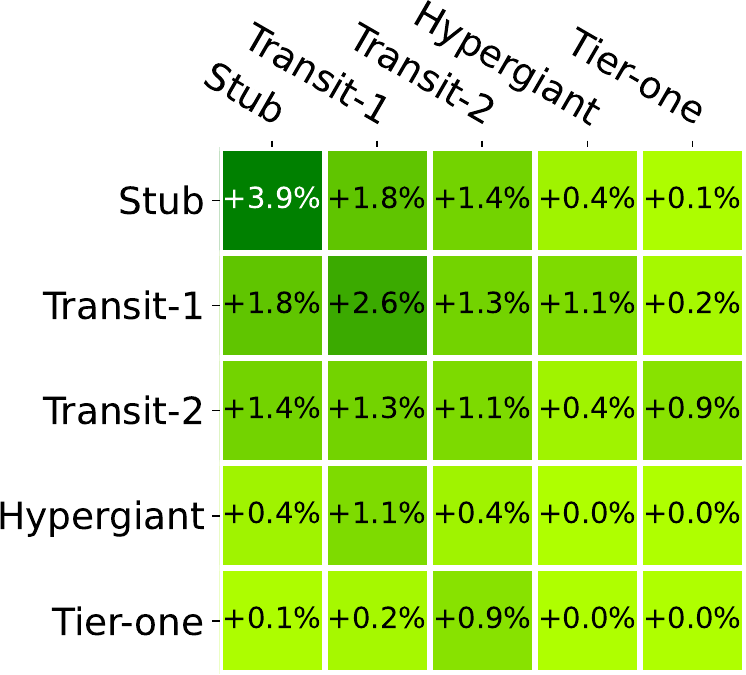}
            \captionsetup{width=1\textwidth}
            \captionof{figure}{\name enables mapping more links than \rname.}
            \label{fig:random_sampling}
        \end{minipage}
        \hspace{0.3cm}
        \begin{minipage}{0.45\textwidth}
            \centering
            \resizebox{4cm}{!}{
                \begin{tabular}{c|cccc}
                    \Xhline{1.5pt}
                    & $\setminus\lbrace{0,1}\rbrace$ & $\setminus\lbrace{2,3}\rbrace$ & $\setminus\lbrace{4,5}\rbrace$ & $\setminus\lbrace{6,7,8}\rbrace$\\\Xhline{1.5pt}\vspace{0.0cm}
                    \First & \ApplyGradient{1.04} & \ApplyGradient{1.05} & \ApplyGradient{1.07} & \ApplyGradient{1.06}\\\cline{2-5}\vspace{0.0cm}
                    \Second & \ApplyGradient{1.17} & \ApplyGradient{1.02} & \ApplyGradient{1.07} & \ApplyGradient{1.34}\\\cline{2-5}\vspace{0.0cm}
                    \Third & \ApplyGradient{1.09} & \ApplyGradient{1.11} & \ApplyGradient{1.09} & \ApplyGradient{1.12}\\\cline{2-5}\vspace{0.0cm}
                    \Fourth & \ApplyGradient{1.32} & \ApplyGradient{1.25} & \ApplyGradient{1.15} & \ApplyGradient{1.16}\\\cline{2-5}\vspace{0.0cm}
                    \Fifth & \ApplyGradient{1.61} & \ApplyGradient{1.59} & \ApplyGradient{1.71} & \ApplyGradient{1.62}\\\Xhline{1.5pt}
                \end{tabular}
            }
            \captionof{table}{Omitting one feature category reduces the performance of \name for every use case.}
            \label{tab:lessfeat}
        \end{minipage}
    \end{minipage}
\end{figure}

\myitem{The balanced sampling avoids biases in the collected data.}
We implement \rname, a modified version
of \name where new-AS-links events used to compute redundancy scores across VPs
are sampled randomly, \ie using the distribution depicted in \fref{fig:actual_sampling}
(instead of using the balanced sampling in~\S\ref{sec:design_events}).
We compare the performance of
\name and \rname on AS topology mapping. Note that
we observe similar results for other use cases.
We map the AS topology
for May 2023 (following the methodology in \S\ref{sec:eval_as_rel})
using both \name and \rname and with the same volume of data in either case. 
\fref{fig:random_sampling} depicts the proportion of additional AS links
that we can map when using \name compared to when using \rname for every new-AS-link category.
\name always yields better or identical performance than \rname.
The highest difference is when mapping stub-to-stub links
(+3.9\%) or Transit1-to-Transit1 links (+2.6\%).
These two link categories are underrepresented when using a random sampling (see \fref{fig:actual_sampling}),
demonstrating that our balanced sampling scheme mitigates biases.

\myitem{Every feature category is useful.} We implement \name $\setminus \lbrace f_i,..,f_j \rbrace$, 
a modified version of \name where we omit features $\lbrace f_i,..,f_j \rbrace$
when computing redundancy scores, with $i...j$ 
the feature indexes in \tref{tab:descfeatures}. We use four different versions of \name 
$\setminus \lbrace ... \rbrace$, each omitting a different feature category.
We show the \spfa of \name over each \name $\setminus \lbrace ... \rbrace$ for use cases \First, \Second, \Third,
\Fourth, and \Fifth in \S\ref{sec:eval_performance},
with the objective of detecting 70\% of the events or mapping 70\% of the
AS-level topology. Regardless of which feature category is omitted, \name performs better (\ie
the \spfa is above 1). We conclude that every feature category is valuable.

%% file: related_work.tex
\section{Related work}

\myitem{Redundancy and bias between the VPs.}
Chen \etal showed that VPs observe identical (redundant) AS links
and that it is possible to reduce the number
of VPs while providing similar measurement power~\cite{redundancy_analysis}.
However, they only focus on one objective (observing AS links) whereas
\name works for \textit{any} objective.
Previous works reported that the VPs are biased
(in terms of location, network size, etc.)~\cite{RIS_pavlos,qualityIBGP, ShavittBias}.
\name is data-driven and does not consider these biases as we show 
that an \textit{unbiased} selection strategy performs poorly (\S\ref{sec:eval_performance}). 

\myitem{Strategies to select VPs.}
Prior works demonstrated that carefully selecting VPs 
increases the utility of the data~\cite{greedy},
and proposed a greedy selection strategy
that performs better than other naive 
approaches~\cite{placing, greedy}.
However, their selection strategy optimizes one objective
(discovering AS links)
and thus lacks generality (\S\ref{sec:eval_performance}).
Recent works also study the impact of the VP selection
on the discovered IP space and AS links \cite{theseAllemande}.

\myitem{Placement of the VPs.}
Gregori \etal proposed a methodology that finds a
relevant placement for a new VP~\cite{Enrico2012}. Roughan \etal
estimated that 700 strategically positioned VPs were 
enough to monitor the Internet
topology~\cite{bigfoot}. Finally, Cittadini \etal demonstrated 
the marginal utility of adding
new VPs at the core of the Internet~\cite{qualityIBGP}.

\myitem{Strategies to select active measurement probes.}
Active measurement platforms (\eg RIPE Atlas) also generate
a large volume of data
and several data-driven approaches for probe selection
exist~\cite{metis,10.1145/3106328.3106334, v6measuresVP}. Unlike \name, these approaches 
optimize the probe selection for specific use cases. %

\myitem{Uses of topological features.}
Previous works computed topological features on the AS topology 
to detect routing anomalies~\cite{9912989,dfoh,ashegemony_poster2017}.

%% file: conclusion.tex
\section{Conclusion}

We uncovered redundancy in the BGP routes exported by the \ris and \rv VPs 
and identified this redundancy as an opportunity
to optimize the use of these data collection systems.
We presented \name, a system that samples BGP data at the VP granularity,
enabling users to improve the coverage and accuracy of their studies without processing more data.

The principles that \name embodies can also lead to a better understanding of the structure
of the global Internet as well as how to optimize the measurement and analysis of its routing system.
For instance, our redundancy scores 
could lead to more strategic approaches to gathering and retaining BGP data,
\eg \ris and \rv could deprioritize VPs which are overwhelmingly redundant with many others, on a more scientific basis.
Finally, our approach can be adapted to active measurement platforms 
(\eg Atlas~\cite{RIPE_Atlas}) to reach the same objective of extensive 
coverage with reduced redundant data.

%% file: ack.tex
\section{Acknowledgements}
This work was supported by the ArtIC project (grant ANR-20-THIA-0006-01), Région Grand Est,
Inria Nancy-Grand Est, IHU of Strasbourg, University of Strasbourg,
University of Haute-Alsace, the RIPE NCC Community Projects Fund, NSF CNS-2120399
and NSF OAC-2131987. Views are those of the authors and
do not represent the endorsements of the funding agencies.

%% file: survey.tex
\section{Survey}
\label{sec:survey}

\myitem{Detailed methdology.}
We selected eleven papers and classified them based on 
how authors collected the BGP data (categories $C_1$ and $C_2$). We then emailed the authors
and asked them about their experience with using BGP routes from \ris and \rv.
We did not have answers for three papers.
We promised to share the answers of the participants in an anonymized fashion.
Thus, we do not show parts of a few answers that would make de-anonymization
possible.  However, the missing parts never change the main message
conveyed in the answers.

\myitem{Detailed answers.}
\tref{tab:survey} lists the questions we asked
the participants of our survey along with their detailed answers.
We color the answers based on whether they are
in favor (green) of using a tool such as \name or not (red).
Neutral answers are colored in blue.
The vast majority of the answers indicate that \name would be beneficial for users
and improve the quality of their measurement studies.

\begin{table*}[ht]
    \renewcommand{\arraystretch}{0.8} %

    \resizebox{18cm}{!}
    {
        \begin{tabular}{@{}cll@{}}
            \toprule
            \multicolumn{1}{l}{\textbf{Collection strategy}}                                                       & \textbf{Questions asked}                                                                                                                                     & \textbf{Collected answers}                                                                                                                                                                                                                                                                                                                                                                                                                                                                                                                                                                                                                                                                                                                                                                                \\ \midrule
            \multirow{4}{*}{\textbf{\begin{tabular}[c]{@{}c@{}} \\ \\ \\ \\ \\ \\ $C_1$: All routes and\\ subset of VPs \\ (seven papers) \end{tabular}}}                   & Why did you use a subset of the VPs ?                                                                                                                        & \begin{tabular}[c]{@{}l@{}} \textbf{\textcolor{green1}{To speed up data processing (x2)}} \\ \textbf{\textcolor{green1}{For disk space and time efficiency (x1)}} \\ \textbf{\textcolor{green1}{I thought the rest would be similar (x1)}} \\ \textcolor{green1}{I did not manage to use them all (x2)} \end{tabular}                                                                                                                                                                                                                                                                                                                                                                                                                                                                                                                                                                                                       \\ \cmidrule(l){2-3} 
                                                                                                                                                                & How did you select your VPs ?                                                                                                                                & \begin{tabular}[c]{@{}l@{}} \textbf{\textcolor{green1}{I took them randomly (x2)}} \\ \textbf{\textcolor{green1}{I do not remember (x2)}} \\ \textcolor{green2}{It was arbitrary: my script partially failed (x1)} \\ \textcolor{green2}{I took geographically distant BGP collectors (x1)} \\ \textcolor{blue1}{I did not manage to use VPs from one data provider (x1)}\end{tabular}                                                                                                                                                                                                                                                                                                                                                                                                                                                                                                                                               \\ \cmidrule(l){2-3} 
                                                                                                                                                                & \begin{tabular}[c]{@{}l@{}}Do you think more VPs would improve\\ the quality of your results?\end{tabular}                                                   & \begin{tabular}[c]{@{}l@{}} \textbf{\textcolor{green1}{Yes (x4)}}\\ \textcolor{green1}{Results would be similar, but it can help to find corner cases (x1)}\\ \textcolor{green1}{Yes, but not significantly (x1)} \\ \textcolor{blue1}{I am not sure (x1)}\end{tabular}                                                                                                                                                                                                                                                                                                                                                                                                                                                                                                                                                                                                                         \\ \cmidrule(l){2-3} 
                                                                                                                                                                & \begin{tabular}[c]{@{}l@{}}Would you have used more VPs\\ if you could?\end{tabular}                                                                         & \begin{tabular}[c]{@{}l@{}} \textbf{\textcolor{green1}{Yes (x4)}}\\ \textbf{\textcolor{green1}{Yes, I'd love to (x1)}}\\ \textbf{\textcolor{green1}{Definitely (x1)}}\\ \textcolor{red}{I am not sure, but I don’t think so (x1)}\end{tabular}                                                                                                                                                                                                                                                                                                                                                                                                                                                                                                                                                                                                                                                                     \\ \midrule
            \multirow{3}{*}{\textbf{\begin{tabular}[c]{@{}c@{}}\\ \\ \\ $C_2$: Limited duration\\ of experiment \\ (five papers) \end{tabular}}}                 & \begin{tabular}[c]{@{}l@{}}Was the processing time a factor\\ that you considered when you decided\\ on the duration of your measurement study?\end{tabular} & \textbf{\textcolor{green1}{Yes (x3)}}                                                                                                                                                                                                                                                                                                                                                                                                                                                                                                                                                                                                                                                                                                                                                                                                  \\ \cmidrule(l){2-3} 
                                                                                                                                                                & \begin{tabular}[c]{@{}l@{}}Do you think extending the duration\\ of your measurement study would\\ improve the quality of your results?\end{tabular}         & \begin{tabular}[c]{@{}l@{}}\textbf{\textcolor{green1}{Yes (x2)}}\\ \textbf{\textcolor{green1}{Yes, especially for rare events (x1)}} \\ \textcolor{green1}{Potentially (x1)}\\ \textcolor{green1}{Yes, but not significantly (x1)}\end{tabular}                                                                                                                                                                                                                                                                                                                                                                                                                                                                                                                                                                                                                                                              \\ \cmidrule(l){2-3} 
                                                                                                                                                                & \begin{tabular}[c]{@{}l@{}}Would have extended the duration\\ of your measurement study\\ if you had more resources?\end{tabular}                            & \begin{tabular}[c]{@{}l@{}}\textbf{\textcolor{green1}{Yes (x2)}}\\ \textcolor{green1}{Yes, but it depends on the time remaining before the deadline (x1)}\\ \textcolor{green1}{I think so, but also if I had more time before the deadline (x1)}\end{tabular}                                                                                                                                                                                                                                                                                                                                                                                                                                                                                                                                                                                                                  \\ \midrule  \midrule
            \multirow{2}{*}{\textbf{\begin{tabular}[c]{@{}c@{}}\\ \\ \\ \\ All eight papers \end{tabular}}}                                                                                  & \begin{tabular}[c]{@{}l@{}}Do you find the data from RIS and\\ RouteViews expensive to process\\ in terms of computational resources?\end{tabular}           & \begin{tabular}[c]{@{}l@{}}\textbf{\textcolor{green1}{Yes (x1)}}\\ \textbf{\textcolor{green1}{Yes, CPU and storage (x2)}}\\ \textbf{\textcolor{green1}{Yes, the storage cost and the download cost are very large (x1)}}\\ \textbf{\textcolor{green1}{CPU is the main issue (x1)}}\\ \textbf{\textcolor{green1}{RIS data takes a lot of time to download, especially when we need data for multiple days (x1)}}\\ \textcolor{green1}{Not the worst, but we definitely need a resourceful server if we want to catch some deadline (x1)}\\ \textcolor{red}{We did that in a server so that was not a huge issue (x1)}\\ \textbf{\textcolor{red}{No (x1)}} \end{tabular}                                                                                                                                                                                                                                                                                                                                                     \\ \cmidrule(l){2-3} 
                                                                                                                                                                & \begin{tabular}[c]{@{}l@{}}Is there any additional challenge\\ that you encountered when processing\\ the BGP data from RIS and RouteViews?\end{tabular}     & \begin{tabular}[c]{@{}l@{}}\textbf{\textcolor{green1}{Our team used Spark clusters and Python but it was too slow (x1)}}\\ \textbf{\textcolor{green1}{We had to download the data from all VPs as there is no optimal solution for selecting them,}}\\  \textbf{\textcolor{green1}{~~~~the storage overhead and time overhead were extremely high (x1)}} \\ \textbf{\textcolor{green1}{It'll be helpful to make processing faster and less resource-consuming (x1)}}\\ \textcolor{green1}{Too many duplicate announcements make processing harder (x1)}\\ \textcolor{green1}{Variable sizes of update files exacerbate scheduling parallelization (x1)} \\ \textcolor{green1}{RIS took a lot longer than RouteViews (x1)} \\ \textcolor{blue1}{We had issues when collecting updates in real-time (x1)} \\ \textcolor{blue1}{We had to deal with bugs in BGPdump (x1)}\\ \textcolor{blue1}{Broken data feeds and data cleanup is also an issue that we need to take care of (x1)} \\ \textcolor{blue1}{Our study was done pre-BGPStream, which would have helped quite a bit already (x1)} \end{tabular}
        \end{tabular}
    }
    \caption{An exhaustive list of the questions asked to the participants of the survey along with their detailed answers.
    We color an answer in (bold) green if it (strongly) motivates the usage of a tool such as \name. Blue answers are neutral, \ie they do not motivate \name but also do not disincentive it.
    Finally, (bold) red answers (strongly) disincentive the usage of a tool such as \name.}
    \label{tab:survey}

\end{table*}

%% file: eval_extended.tex
\section{Extended evaluation}
\label{app:eval}

In this section, we evaluate the performances of \namevboth
(\tref{tab:benchmark}) and \namevsix  (\tref{tab:benchmark_v6}) on
the five use cases presented in \S\ref{sec:eval_performance}, namely
transient paths detection (\First), MOAS detection (\Second),
AS topology mapping (\Third), traffic engineering detection (\Fourth),
and unnecessary updates detection (\Fifth). Similarly to \S\ref{sec:eval_performance},
we compare \namevsix and \namevboth against the three naive baselines (random, AS-distance, and
unbiased) as well as the eight \textit{greedy specific} VPs selection strategies
(three optimized for Def. \ref{def:red1}, \ref{def:red2}, and \ref{def:red3}
and one optimized for each of the five use cases). We present the results
in terms of data \spfa, as defined in \S\ref{sec:eval_performance}.

\myitem{\namevboth and \namevsix outperform the three naive baselines for every objective.}
For \namevboth, the \spfa can be as high as 6.57 when trying to detect 50\% of
the traffic engineering paths while for \namevsix it
can be as high as 5.05 when trying to map 50\% of the AS topology.
On average, \namevboth only needs 41.6\% of the 
data (\spfa of 2.4) required by a naive baseline to meet the same objective
while \namevsix needs 44.5\% (\spfa of 2.26).

\myitem{\namevboth and \namevsix prevent overfitting.}
For the vast majority of the objectives,
\textit{greedy specific} performs better than \namevboth or \namevsix only for the
use cases for which it is optimized. There are a few cases where \textit{greedy
specific} performs better than \namevboth or \namevsix for a use case that it does not optimized.
For instance, \namevsix needs to process 20\% (\spfa of 0.8) more data than \textit{greedy specific} optimized
for use case \First to detect 90\% of the MOAS (use case \Second).
However, in the vast majority of the cases, both \namevboth and \namevsix outperform the \textit{greedy specifics}.
For instance, \namevsix only needs 26\% (\spfa of 3.74) of the volume required by
the \textit{greedy specific} optimized for use case \Fourth to detect 90\%
of the MOAS (use case \Second). These results show that
\name does not overfit while \textit{greedy specific} does.

\begin{table*}[!t]
	\centering
	\resizebox{15cm}{!}{
	\begin{tabular}{cc|ccc|ccccc|ccc}
		\Xhline{1.5pt}
		 \multirow{2}{*}{Use case} & \multirow{2}{*}{Objective} & \multicolumn{3}{c|}{Naives baselines} & \multicolumn{5}{c|}{Greedy specifics use cases (\S\ref{sec:eval_performance})} & \multicolumn{3}{c}{Greedy specifics Def. (\S\ref{sec:opportunity})}\\\cline{3-6}\cline{7-13}
		 &  & Random & AS-distance & \bb & \First & \Second & \Third & \Fourth & \Fifth & Def. 1 & \hspace{0.3cm}Def. 2 & Def. 3\\\Xhline{1.5pt}\vspace{0.0cm}
		\multirow{4}{*}{\vspace{0.4cm}\fboxrule=0pt\fbox{\parbox[c]{2.1cm}{\small \centering \textbf{Transient path detection\\(\First)}}}} & 50 \% & \ApplyGradient{1.32} & \ApplyGradient{1.87} & \ApplyGradient{1.94} & \ApplyGradient{0.61} & \ApplyGradient{1.19} & \ApplyGradient{1.11} & \ApplyGradient{1.36} & \ApplyGradient{1.21} & \ApplyGradient{2.08} & \ApplyGradient{2.24} & \ApplyGradient{1.79}\\\cline{2-13}\vspace{0.0cm}
		 & 70 \% & \ApplyGradient{1.38} & \ApplyGradient{1.62} & \ApplyGradient{1.83} & \ApplyGradient{0.74} & \ApplyGradient{1.30} & \ApplyGradient{1.15} & \ApplyGradient{1.40} & \ApplyGradient{1.18} & \ApplyGradient{1.78} & \ApplyGradient{1.97} & \ApplyGradient{1.53}\\\cline{2-13}\vspace{0.0cm}
		 & 90 \% & \ApplyGradient{1.16} & \ApplyGradient{1.42} & \ApplyGradient{1.40} & \ApplyGradient{0.71} & \ApplyGradient{1.39} & \ApplyGradient{1.74} & \ApplyGradient{1.69} & \ApplyGradient{1.21} & \ApplyGradient{1.34} & \ApplyGradient{1.36} & \ApplyGradient{1.40}\\\Xhline{1.5pt}\vspace{0.0cm}
		\multirow{4}{*}{\vspace{0.4cm}\fboxrule=0pt\fbox{\parbox[c]{2.1cm}{\small \centering \textbf{MOAS detection\\(\Second)}}}} & 50 \% & \ApplyGradient{1.93} & \ApplyGradient{3.38} & \ApplyGradient{4.03} & \ApplyGradient{1.95} & \ApplyGradient{0.78} & \ApplyGradient{1.41} & \ApplyGradient{2.05} & \ApplyGradient{1.37} & \ApplyGradient{3.34} & \ApplyGradient{3.21} & \ApplyGradient{2.88}\\\cline{2-13}\vspace{0.0cm}
		 & 70 \% & \ApplyGradient{1.96} & \ApplyGradient{3.49} & \ApplyGradient{4.16} & \ApplyGradient{2.14} & \ApplyGradient{0.68} & \ApplyGradient{1.91} & \ApplyGradient{2.52} & \ApplyGradient{1.56} & \ApplyGradient{2.91} & \ApplyGradient{2.60} & \ApplyGradient{2.81}\\\cline{2-13}\vspace{0.0cm}
		 & 90 \% & \ApplyGradient{1.16} & \ApplyGradient{1.69} & \ApplyGradient{2.07} & \ApplyGradient{1.52} & \ApplyGradient{0.69} & \ApplyGradient{1.68} & \ApplyGradient{1.87} & \ApplyGradient{1.53} & \ApplyGradient{1.31} & \ApplyGradient{1.25} & \ApplyGradient{1.40}\\\Xhline{1.5pt}\vspace{0.0cm}
		\multirow{4}{*}{\vspace{0.4cm}\fboxrule=0pt\fbox{\parbox[c]{2.1cm}{\small \centering \textbf{AS topology mapping\\(\Third)}}}} & 50 \% & \ApplyGradient{2.47} & \ApplyGradient{2.90} & \ApplyGradient{2.72} & \ApplyGradient{1.18} & \ApplyGradient{1.02} & \ApplyGradient{0.58} & \ApplyGradient{1.45} & \ApplyGradient{1.41} & \ApplyGradient{2.38} & \ApplyGradient{2.30} & \ApplyGradient{2.16}\\\cline{2-13}\vspace{0.0cm}
		 & 70 \% & \ApplyGradient{2.27} & \ApplyGradient{2.52} & \ApplyGradient{2.29} & \ApplyGradient{1.26} & \ApplyGradient{1.14} & \ApplyGradient{0.68} & \ApplyGradient{1.25} & \ApplyGradient{1.19} & \ApplyGradient{2.03} & \ApplyGradient{1.71} & \ApplyGradient{2.03}\\\cline{2-13}\vspace{0.0cm}
		 & 90 \% & \ApplyGradient{1.71} & \ApplyGradient{1.85} & \ApplyGradient{1.78} & \ApplyGradient{1.14} & \ApplyGradient{1.13} & \ApplyGradient{0.82} & \ApplyGradient{1.17} & \ApplyGradient{1.15} & \ApplyGradient{1.62} & \ApplyGradient{1.61} & \ApplyGradient{1.56}\\\Xhline{1.5pt}\vspace{0.0cm}
		\multirow{4}{*}{\vspace{0.4cm}\fboxrule=0pt\fbox{\parbox[c]{2.1cm}{\small \centering \textbf{Traffic engineering detection\\(\Fourth)}}}} & 50 \% & \ApplyGradient{3.77} & \ApplyGradient{6.57} & \ApplyGradient{4.43} & \ApplyGradient{3.21} & \ApplyGradient{1.89} & \ApplyGradient{1.47} & \ApplyGradient{0.47} & \ApplyGradient{2.57} & \ApplyGradient{3.43} & \ApplyGradient{2.89} & \ApplyGradient{2.57}\\\cline{2-13}\vspace{0.0cm}
		 & 70 \% & \ApplyGradient{2.34} & \ApplyGradient{3.17} & \ApplyGradient{2.56} & \ApplyGradient{2.20} & \ApplyGradient{1.60} & \ApplyGradient{1.93} & \ApplyGradient{0.35} & \ApplyGradient{2.06} & \ApplyGradient{1.97} & \ApplyGradient{2.01} & \ApplyGradient{2.05}\\\cline{2-13}\vspace{0.0cm}
		 & 90 \% & \ApplyGradient{1.90} & \ApplyGradient{2.02} & \ApplyGradient{1.76} & \ApplyGradient{2.02} & \ApplyGradient{1.78} & \ApplyGradient{1.94} & \ApplyGradient{0.31} & \ApplyGradient{1.93} & \ApplyGradient{2.13} & \ApplyGradient{1.67} & \ApplyGradient{1.93}\\\Xhline{1.5pt}\vspace{0.0cm}
		\multirow{4}{*}{\vspace{0.4cm}\fboxrule=0pt\fbox{\parbox[c]{2.1cm}{\small \centering \textbf{Unnecessary updates detection\\(\Fifth)}}}} & 50 \% & \ApplyGradient{2.13} & \ApplyGradient{4.10} & \ApplyGradient{3.15} & \ApplyGradient{2.41} & \ApplyGradient{2.54} & \ApplyGradient{2.72} & \ApplyGradient{3.12} & \ApplyGradient{0.41} & \ApplyGradient{2.94} & \ApplyGradient{2.78} & \ApplyGradient{2.83}\\\cline{2-13}\vspace{0.0cm}
		 & 70 \% & \ApplyGradient{1.27} & \ApplyGradient{1.95} & \ApplyGradient{1.80} & \ApplyGradient{1.47} & \ApplyGradient{1.12} & \ApplyGradient{1.28} & \ApplyGradient{1.59} & \ApplyGradient{0.35} & \ApplyGradient{1.71} & \ApplyGradient{1.81} & \ApplyGradient{1.57}\\\cline{2-13}\vspace{0.0cm}
		 & 90 \% & \ApplyGradient{1.01} & \ApplyGradient{1.29} & \ApplyGradient{1.35} & \ApplyGradient{1.04} & \ApplyGradient{0.85} & \ApplyGradient{0.96} & \ApplyGradient{1.09} & \ApplyGradient{0.46} & \ApplyGradient{1.00} & \ApplyGradient{1.11} & \ApplyGradient{1.11}\\\Xhline{1.5pt}\vspace{0.0cm}
	\end{tabular}
	}
	\caption{Data reduction factor for \namevboth compared to several baselines for five use cases. 
    \namevboth enables to detect 70\% of the MOAS using only 28.6\% (\spfa of 3.49) of the volume required by
    the AS distance baseline to meet the same objective. The average reduction factor over all objectives and naive 
    baselines is 2.25.}
	\label{tab:benchmark}
\end{table*}

\begin{table*}[!t]
	\centering
	\resizebox{15cm}{!}{
	\begin{tabular}{cc|ccc|ccccc|ccc}
		\Xhline{1.5pt}
		 \multirow{2}{*}{Use case} & \multirow{2}{*}{Objective} & \multicolumn{3}{c|}{Naives baselines} & \multicolumn{5}{c|}{Greedy specifics use cases (\S\ref{sec:eval_performance})} & \multicolumn{3}{c}{Greedy specifics Def. (\S\ref{sec:opportunity})}\\\cline{3-6}\cline{7-13}
		 &  & Random & AS-distance & \bb & \First & \Second & \Third & \Fourth & \Fifth & Def. 1 & \hspace{0.3cm}Def. 2 & Def. 3\\\Xhline{1.5pt}\vspace{0.0cm}
		\multirow{4}{*}{\vspace{0.4cm}\fboxrule=0pt\fbox{\parbox[c]{2.1cm}{\small \centering \textbf{Transient path detection\\(\First)}}}} & 50 \% & \ApplyGradient{1.43} & \ApplyGradient{1.65} & \ApplyGradient{2.29} & \ApplyGradient{0.44} & \ApplyGradient{1.11} & \ApplyGradient{1.14} & \ApplyGradient{1.67} & \ApplyGradient{1.20} & \ApplyGradient{1.56} & \ApplyGradient{1.36} & \ApplyGradient{1.90}\\\cline{2-13}\vspace{0.0cm}
		 & 70 \% & \ApplyGradient{1.71} & \ApplyGradient{1.84} & \ApplyGradient{2.00} & \ApplyGradient{0.64} & \ApplyGradient{1.59} & \ApplyGradient{1.96} & \ApplyGradient{2.88} & \ApplyGradient{2.25} & \ApplyGradient{1.75} & \ApplyGradient{1.86} & \ApplyGradient{1.67}\\\cline{2-13}\vspace{0.0cm}
		 & 90 \% & \ApplyGradient{1.52} & \ApplyGradient{1.43} & \ApplyGradient{1.42} & \ApplyGradient{0.62} & \ApplyGradient{1.49} & \ApplyGradient{1.49} & \ApplyGradient{1.79} & \ApplyGradient{1.48} & \ApplyGradient{1.51} & \ApplyGradient{1.72} & \ApplyGradient{2.11}\\\Xhline{1.5pt}\vspace{0.0cm}
		\multirow{4}{*}{\vspace{0.4cm}\fboxrule=0pt\fbox{\parbox[c]{2.1cm}{\small \centering \textbf{MOAS detection\\(\Second)}}}} & 50 \% & \ApplyGradient{1.94} & \ApplyGradient{1.65} & \ApplyGradient{2.37} & \ApplyGradient{1.10} & \ApplyGradient{0.21} & \ApplyGradient{0.36} & \ApplyGradient{1.56} & \ApplyGradient{2.33} & \ApplyGradient{1.04} & \ApplyGradient{0.73} & \ApplyGradient{1.30}\\\cline{2-13}\vspace{0.0cm}
		 & 70 \% & \ApplyGradient{4.24} & \ApplyGradient{1.70} & \ApplyGradient{3.25} & \ApplyGradient{1.26} & \ApplyGradient{0.51} & \ApplyGradient{1.05} & \ApplyGradient{4.98} & \ApplyGradient{3.38} & \ApplyGradient{4.20} & \ApplyGradient{3.71} & \ApplyGradient{4.13}\\\cline{2-13}\vspace{0.0cm}
		 & 90 \% & \ApplyGradient{3.03} & \ApplyGradient{1.75} & \ApplyGradient{2.19} & \ApplyGradient{0.80} & \ApplyGradient{0.53} & \ApplyGradient{2.67} & \ApplyGradient{3.74} & \ApplyGradient{2.56} & \ApplyGradient{3.54} & \ApplyGradient{3.69} & \ApplyGradient{3.84}\\\Xhline{1.5pt}\vspace{0.0cm}
		\multirow{4}{*}{\vspace{0.4cm}\fboxrule=0pt\fbox{\parbox[c]{2.1cm}{\small \centering \textbf{AS topology mapping\\(\Third)}}}} & 50 \% & \ApplyGradient{4.45} & \ApplyGradient{3.68} & \ApplyGradient{5.05} & \ApplyGradient{1.49} & \ApplyGradient{0.72} & \ApplyGradient{0.54} & \ApplyGradient{2.41} & \ApplyGradient{3.03} & \ApplyGradient{1.92} & \ApplyGradient{1.65} & \ApplyGradient{3.29}\\\cline{2-13}\vspace{0.0cm}
		 & 70 \% & \ApplyGradient{2.83} & \ApplyGradient{3.26} & \ApplyGradient{3.14} & \ApplyGradient{1.18} & \ApplyGradient{1.14} & \ApplyGradient{0.73} & \ApplyGradient{2.07} & \ApplyGradient{1.38} & \ApplyGradient{2.27} & \ApplyGradient{2.17} & \ApplyGradient{2.48}\\\cline{2-13}\vspace{0.0cm}
		 & 90 \% & \ApplyGradient{1.86} & \ApplyGradient{2.00} & \ApplyGradient{1.99} & \ApplyGradient{1.10} & \ApplyGradient{1.12} & \ApplyGradient{0.86} & \ApplyGradient{1.25} & \ApplyGradient{1.30} & \ApplyGradient{1.56} & \ApplyGradient{1.70} & \ApplyGradient{2.02}\\\Xhline{1.5pt}\vspace{0.0cm}
		\multirow{4}{*}{\vspace{0.4cm}\fboxrule=0pt\fbox{\parbox[c]{2.1cm}{\small \centering \textbf{Traffic engineering detection\\(\Fourth)}}}} & 50 \% & \ApplyGradient{2.27} & \ApplyGradient{1.68} & \ApplyGradient{1.34} & \ApplyGradient{2.68} & \ApplyGradient{0.51} & \ApplyGradient{0.58} & \ApplyGradient{0.12} & \ApplyGradient{1.89} & \ApplyGradient{0.75} & \ApplyGradient{0.95} & \ApplyGradient{0.53}\\\cline{2-13}\vspace{0.0cm}
		 & 70 \% & \ApplyGradient{3.76} & \ApplyGradient{5.14} & \ApplyGradient{2.86} & \ApplyGradient{3.03} & \ApplyGradient{2.64} & \ApplyGradient{3.03} & \ApplyGradient{0.30} & \ApplyGradient{4.66} & \ApplyGradient{2.07} & \ApplyGradient{1.61} & \ApplyGradient{1.14}\\\cline{2-13}\vspace{0.0cm}
		 & 90 \% & \ApplyGradient{1.29} & \ApplyGradient{1.36} & \ApplyGradient{1.18} & \ApplyGradient{1.49} & \ApplyGradient{1.49} & \ApplyGradient{1.49} & \ApplyGradient{0.65} & \ApplyGradient{1.49} & \ApplyGradient{0.88} & \ApplyGradient{0.56} & \ApplyGradient{1.19}\\\Xhline{1.5pt}\vspace{0.0cm}
		\multirow{4}{*}{\vspace{0.4cm}\fboxrule=0pt\fbox{\parbox[c]{2.1cm}{\small \centering \textbf{Unnecessary updates detection\\(\Fifth)}}}} & 50 \% & \ApplyGradient{1.45} & \ApplyGradient{2.19} & \ApplyGradient{2.63} & \ApplyGradient{1.57} & \ApplyGradient{2.94} & \ApplyGradient{1.97} & \ApplyGradient{3.21} & \ApplyGradient{0.22} & \ApplyGradient{2.44} & \ApplyGradient{2.70} & \ApplyGradient{1.95}\\\cline{2-13}\vspace{0.0cm}
		 & 70 \% & \ApplyGradient{1.37} & \ApplyGradient{2.13} & \ApplyGradient{2.09} & \ApplyGradient{1.79} & \ApplyGradient{1.98} & \ApplyGradient{2.07} & \ApplyGradient{2.26} & \ApplyGradient{0.31} & \ApplyGradient{1.91} & \ApplyGradient{2.18} & \ApplyGradient{1.82}\\\cline{2-13}\vspace{0.0cm}
		 & 90 \% & \ApplyGradient{1.23} & \ApplyGradient{1.46} & \ApplyGradient{1.58} & \ApplyGradient{1.34} & \ApplyGradient{1.39} & \ApplyGradient{1.46} & \ApplyGradient{1.69} & \ApplyGradient{0.50} & \ApplyGradient{1.83} & \ApplyGradient{1.49} & \ApplyGradient{1.45}\\\Xhline{1.5pt}\vspace{0.0cm}
	\end{tabular}
	}
	\caption{Data reduction factor for \namevsix compared to several baselines for five use cases. 
    \namevsix enables to detect 90\% of the MOAS using only 33\% (\spfa of 3.03) of the volume required by
    the random selection to meet the same objective. The average reduction factor over all objectives and naive 
    baselines is 2.25.}
	\label{tab:benchmark_v6}
\end{table*}